\documentclass[journal ]{new-aiaa}
\usepackage[utf8]{inputenc}
\usepackage{textcomp}

\usepackage{graphicx}
\usepackage{amsmath}
\usepackage[version=4]{mhchem}
\usepackage{siunitx}
\usepackage{longtable,tabularx}
\usepackage{comment}
\usepackage{subfigure}
\usepackage{tabularray}
\usepackage{hyperref}
\usepackage{footnpag}
\usepackage{fancyhdr}
\title{Accounting for Solar Radiation Pressure in the Hamiltonian Normal Form of the Elliptic Restricted Three-Body Problem}
\author{Carson Hunsberger \footnote{Graduate Student, Department of Aerospace Engineering, Hammond Building, University Park, PA 16802.}}
\affil{The Pennsylvania State University, University Park, PA, 16802}
\author{David Schwab}
\affil{Air Force Research Lab, Albuquerque, NM, 87116}
\author{Roshan Eapen\footnote{Assistant Professor, Department of Aerospace Engineering, Hammond Building, University Park, PA 16802.}}
\affil{The Pennsylvania State University, University Park, PA, 16802}
\author{Puneet Singla\footnote{Professor, Department of Aerospace Engineering, Hammond Building, University Park, PA 1680.2}}
\affil{The Pennsylvania State University, University Park, PA, 16802}

\newcommand{\xtil}{\tilde{x}}

\newcommand{\pxtil}{\tilde{p}_x}

\begin{document}

\maketitle

\begin{abstract}
Hamiltonian normal forms allow for the analytical approximation of center manifold trajectories and their invariant manifolds through the separation of the saddle and center subspaces that make up the dynamics at the collinear libration points within the elliptic restricted three-body problem. The circular restricted three-body problem is a special case of the elliptic problem-- one that does not take into account the eccentricity of the true orbits of the primaries and thus provides a dynamical model of varying accuracy depending on the true anomaly of the primaries. This paper first shows that the normal forms of the elliptic problem offer nearly identical trajectory characterization capabilities to those of the circular problem and then demonstrates the difference in fidelity by comparing the circular and elliptic normal form representations of ephemeris data for the James Webb Space Telescope. Furthermore, methodology for including solar radiation pressure within the normal form is introduced, and the same ephemeris data is used to demonstrate the resulting increase in fidelity of the dynamical model.
  
\end{abstract}

\section{Introduction}
\thispagestyle{fancy}
\fancyhf{}
\fancyfoot[C]{DISTRIBUTION A: Approved for public release; distribution is unlimited. Public Affairs release approval AFRL-2025-5110.}
In a system with two massive primaries orbiting their shared barycenter, there are particular locations at which the gravitational forces acting on a hypothetical satellite balance out in such a way that the satellite's motion with respect to the barycenter matches that of the primaries. Such points are called libration points, and they have been increasingly viewed as desirable locations for missions. The neighborhoods around the libration points exhibit rich dynamics with several families of quasiperiodic trajectories-- trajectories that exhibit bounded, predictable motion. In the Sun-Earth system, these libration points exist approximately one million kilometers from the Earth, making them the ideal location for observatories such as SOHO, Gaia, and the James Webb Space Telescope, far from the interference of Earth's atmosphere. In the Earth-Moon system, there are libration points both between the Earth and the Moon and on the far side of the Moon, providing locations for potential relays for communication with Earth as well as access to the Moon itself.

Many low-fidelity models approximate such a three-body system. The one that is the most extensively studied is the circular restricted three-body problem (CR3BP), which assumes that both primaries orbit their barycenter in a circular manner. The frequently used synodic frame rotates at a constant rate that matches the orbits of the primaries such that the primaries remain at fixed locations within the frame. The CR3BP has five libration points that can, in turn, be separated into two groups based on their dynamical structure. Two libration points, $L_4$ and $L_5$, are often referred to as the triangular points, and are located at the vertices of the two equilateral triangles that can be formed with one side connecting the two primaries. These points have a center$\times$center$\times$center structure, meaning that the linearized dynamics of the triangular points are marginally stable. The other three libration points lie on the line connecting the two primaries, and are often referred to as the collinear points. The collinear points have a saddle$\times$center$\times$center structure, meaning that they are unstable.

There are a few dynamical problems designed specifically to model the Earth-Moon system. These include the bicircular four-body restricted problem (BCP or BCR4BP), which assumes that there is a third massive primary which also orbits the barycenter in a circular manner, the quasi-bicircular problem (QBCP), in which the three primaries orbit in a near-circular coherent manner, and the Hill restricted four-body problem (HR4BP), which assumes that the third primary is sufficiently far from the other two. While they are higher-fidelity models for the Earth-Moon system, they are limited in the sense that they cannot be applied to many other systems. 

The elliptic restricted three-body problem (ER3BP) is a low-fidelity model for any celestial system consisting of two massive primaries orbiting each other in an elliptical manner. Similar to the CR3BP, a frame is defined such that it rotates in a manner that matches the rotation of the primaries. Within this frame, the primaries appear to oscillate along the $x$-axis with an amplitude proportional to the eccentricity of the system. To remove this oscillation, the frame is made to pulsate such that the distance between the primaries is always unitary, resulting in the primaries residing at fixed locations within the rotating-pulsating frame. The libration points of the ER3BP are located at the same fixed positions within the rotating-pulsating frame as those of the CR3BP, and they have the same dynamical properties. Indeed, the CR3BP is merely a special case of the ER3BP-- one in which the eccentricity of the orbits of the primaries is zero. 

In all of these systems, it can be difficult or tedious to numerically obtain quasiperiodic trajectories and their invariant stable and unstable manifolds. Hamiltonian normal forms provide analytical approximations of several families of center manifold trajectories near the libration points, along with their invariant manifolds. The normal forms approximate the dynamical behavior of an entire neighborhood, rather than a single trajectory or manifold. Additionally, they provide constants of motion near the libration points, and have even been referred to as a ``local orbital element equivalent'' \cite{peterson2023local} due to the intuitive trajectory paramaterization that these constants of motion provide.

The normal forms of the CR3BP have been studied extensively \cite{jorba1999methodology,schwab2024characterizing,zhao2025lie,peterson2023local}, and more recently have been leveraged for optimal control purposes \cite{schwabCislunarTransportCharacterization2024,hunsbergerschwab2025hawaii}. The normal forms of the BCP have also been derived \cite{jorba2020vicinity,gomez2001dynamics}; however, it is not a coherent model and also lacks a dynamical equivalent of the CR3BP $L_2$ libration point. The QBCP and its normal forms are first introduced in \cite{andreu1998quasi}, and offer a coherent model close to the BCP in addition to a dynamical equivalent of $L_2$. The QCP normal forms are further studied in \cite{andreu1998quasi,andreu2002dynamics,le2017invariant,rosales2020effect,gabern2001restricted}. The HR4BP also offers a coherent model, and its normal form is discussed in \cite{peterson2023vicinity,peterson2024dynamics}. The $L_2$ HR4BP normal form yields similar results to those of the QBCP.

The Birkhoff normal form of the ER3BP is first derived in \cite{paez2021transits}, followed closely by the derivation of the resonant normal form in \cite{paez2022semi}. Unfortunately, the coefficients of the normal form Hamiltonians found were of very large magnitudes, decreasing the radius of validity of the normal forms. What this meant was that the normal forms were valid only extremely close to the libration points. However, it is demonstrated in \cite{celletti2024dynamics} that a resonant normal form Hamiltonian with a much larger radius of validity can be found.

This paper first provides the full derivations of both the Birkhoff and Resonant normal forms of the ER3BP and demonstrates that both offer reasonable radii of validity around the $L_1$ and $L_2$ libration points. Next, it is shown how the ER3BP normal forms can be used to characterize trajectories near the libration points. Not only can several families of center manifold trajectories be parameterized in a straightforward way, but also their stable and unstable manifolds. Furthermore, it is shown that the ER3BP normal forms offer the same long-term characterization of unstable trajectories as the CR3BP normal forms through analysis similar to that which was conducted in \cite{schwab2024characterizing}. This paper then introduces the methodology for including a simple cannonball model for solar radiation pressure (SRP) for a particular satellite within the normal form itself, allowing for a more accurate dynamical model for systems involving the Sun. The higher fidelity of the ER3BP normal forms is then demonstrated by taking ephemeris data of the James Webb Space Telescope (JWST), as well as a few other missions residing near the Sun-Earth libration points and transforming it into the normal form and action angle spaces for both the CR3BP and ER3BP. While the ER3BP on its own is a noticeable improvement upon the CR3BP, the addition of SRP to the ER3BP normal form yields an even higher fidelity model of the true system. The results included in this paper justify the leveraging of Hamiltonian normal forms for trajectory characterization and optimal control purposes.
\section{The Elliptic Restricted Three-Body Problem}
In the ER3BP, there are two massive primaries in planar elliptic orbits about their barycenter, with a massless satellite present. The ER3BP can be viewed as a more general case of the CR3BP and depends on two parameters. The mass parameter of the system, $\mu$, is defined in an identical manner to the CR3BP-- it is the ratio between the mass of the smaller primary and the total mass of the system. Let $m_1$ be the mass of the larger primary and $m_2$ the mass of the smaller primary, then
\begin{equation}
\mu = \frac{m_2}{m_1+m_2}.
\end{equation}
Unlike in the circular case, the distance between the primaries will generally be changing over time. Let $r$ denote the distance between the primaries, then
\begin{equation}
r = \frac{a(1-e^2)}{1+e\cos f},
\end{equation}
where $a$ is the semimajor axis, $e$ is the eccentricity, and $f$ is the true anomaly \cite{szebehely2012theory}. 

The primaries are at the usual fixed positions on the x-axis within a rotating-pulsating frame ($\mathbf r_1 = [-\mu,0,0]^T,\,\mathbf r_2 = [1-\mu,0,0]^T$).
The equations of motion of the ER3BP with true anomaly as the independent variable are \cite{szebehely2012theory}:
\begin{equation}
\begin{aligned}
x''  &= 2y'+\frac{1}{1+e\cos f}\left(x-\frac{(1-\mu)(x+\mu)}{r_1^3}-\frac{\mu(x+\mu-1)}{r_2^3}\right)\\
y'' &= -2x'+\frac{1}{1+e\cos f}\left(y-\frac{(1-\mu)y}{r_1^3}-\frac{\mu y}{r_2^3}\right)\\
z'' &= -z+\frac{1}{1+e\cos f}\left(z-\frac{(1-\mu)z}{r_1^3}-\frac{\mu z}{r_2^3}\right)
\end{aligned}
\end{equation}
where $r_1=\sqrt{(x+\mu)^2+y^2+z^2)}$ and $r_2 = \sqrt{(x+\mu-1)^2+y^2+z^2}$.
Note that throughout this work, $()'$ will be used to denote $\frac{\mathrm d}{\mathrm d f}$ while $\dot{()}$ will denote $\frac{\mathrm d}{\mathrm d t}$.

The Hamiltonian of the spatial ER3BP is the following \cite{paez2021transits}
\begin{equation}\label{eq:ER3BPHamiltonian}
H = \frac{1}{2}\left(p_x^2+p_y^2+p_z^2\right) - p_yx+p_xy + \frac{1}{1+e\cos f}\left(\frac{1}{2}e\cos f\left(x^2+y^2+z^2 \right)-\frac{1-\mu}{r_1}-\frac{\mu}{r_2}\right),
\end{equation}
where the transformation between the rotating-pulsating state and the Hamiltonian state (consisting of the coordinates $q_j$ and conjugate momenta $p_j$) is given by
\begin{equation}
p_x=x'-y,\quad p_y = y'+x,\,\,\textrm{and}\,\,\,p_z=z'.
\end{equation}
It is clear from \eqref{eq:ER3BPHamiltonian} that the special case where $e=0$ (i.e., a circular orbit), produces the standard CR3BP Hamiltonian as seen in \cite{schwab2024characterizing}.
\section{The Birkhoff Normal Form of the ER3BP}
This section covers all of the steps necessary for obtaining the Birkhoff normal form of the ER3BP about $L_1$ and $L_2$ in a similar manner to \cite{paez2021transits} up until \eqref{eq:genfunc}, where a different equation for the generating functions is used. The last subsection discusses the coordinate transformations necessary for converting between the original state, the normal form state, and the action-angle state. The numerical coordinate transformation introduced in \cite{schwab2024characterizing} is presented as a computationally expensive but generally more accurate alternative to the analytical transformations.
\subsection{Summary of Steps}
  \begin{enumerate}
    \item \hyperref[subsec1]{Expand the Hamiltonian about a libration point.}
    \item \hyperref[subsec2]{Find and apply a symplectic Floquet change of variables $P(f)$ to remove the dependence on true anomaly from the quadratic part of the Hamiltonian.}
    \item \hyperref[subsec3]{Diagonalize the now-autonomous quadratic part of the Hamiltonian.}
    \item \hyperref[subsec4]{Complexify the Hamiltonian.}
    \item \hyperref[subsec5]{Apply a sequence of Lie series transformations to remove the non-resonant terms and terms dependent on true anomaly from the Hamiltonian.}
    \item \hyperref[subsec6]{Realify the Hamiltonian to obtain the real normal form.}
    \item \hyperref[subsec7]{Transform to action-angle variables.}
\end{enumerate}
\subsection{Hamiltonian Expansion}
\label{subsec1}
Similar to the process of obtaining the CR3BP normal form, the potential terms must be expanded as polynomials in the coordinates $q_i$. Unlike in the CR3BP case, however, \eqref{eq:ER3BPHamiltonian} contains an additional scaling factor that must be expanded as a Fourier series.
The full expansion of the Hamiltonian will be a product of the two, yielding a Taylor-Fourier expansion. Before expanding the Hamiltonian, the origin is shifted to the libration point of interest, and the coordinates are rescaled such that the distance between the libration point and the nearest primary is equal to one.
\begin{equation}
\begin{aligned}
x &= \gamma q_1 + (1-\mu)\mp\gamma \\
y &= \gamma q_2\\
z &= \gamma q_3 
\end{aligned} \quad\quad
\begin{aligned}
p_x &= \gamma p_1 \\
p_y &= \gamma p_2 + (1-\mu)\mp\gamma \\
p_z &= \gamma p_3
\end{aligned}
\end{equation}
where $\gamma$ is the distance from the libration point to the nearest primary. The upper sign corresponds to $L_1$ and the lower to $L_2$. The purpose of this transformation is twofold-- shifting the origin to the desired libration point will remove the terms of order one from the Taylor expansion, and the rescaling ensures that the polynomials are well-behaved in the region between the libration point and the nearest primary.
\subsubsection{Taylor Expansion}
The potential term is identical to the one found in the CR3BP, and the same Taylor expansion can be performed leveraging Legendre polynomials, $P_n$, as described in \cite{richardson1980analytic}.
\begin{equation}
\frac{1-\mu}{r_1} + \frac{\mu}{r_2} = \sum_{n=0}^\infty c_n\rho^nP_n\left(\frac{q_1}{\rho}\right)
\end{equation}
Where $\rho = \sqrt{q_1^2+q_2^2+q_3^2}$ and  
\begin{equation}
c_n = \frac{1}{\gamma^3}\left[(\pm1)^n\mu + (-1)^n(1-\mu)\left(\frac{\gamma}{1\mp\gamma}\right)^{n+1}\right].
\end{equation}
Note that the term of degree zero in the expansion can be neglected from the Hamiltonian without affecting the dynamics. Furthermore, since the expansion is being performed about a libration point, the first-degree terms vanish. Therefore, in practice, only the terms of degree $n=2$ to $n=N$ should be included in the Hamiltonian, where $N$ is a user-specified truncation degree.
\subsubsection{Fourier Expansions}
The trigonometric terms can be expanded as real Fourier cosine series in the following manner
\begin{equation}
\frac{1}{1+e\cos f} = \frac{1}{\sqrt{1-e^2}}\left[1+2\sum^\infty_{n=1}\eta^\nu\cos (\nu f)\right]
\end{equation}
\begin{equation}
\frac{e\cos f}{1+e\cos f} = \frac{e}{\sqrt{1-e^2}}\left[\eta+\sum^\infty_{\nu=1}\left(\eta^{\nu+1}+\eta^{\nu-1}\right)\cos(\nu f)\right],
\end{equation}
where $\eta=\frac{1}{e}(\sqrt{1-e^2}-1)$.
We prefer that each term in the Hamiltonian be of the form
\begin{equation*} h^{\mathbf m, \mathbf l}_\nu q_1^{m_1} q_2^{m_2}q_3^{m_3}p_1^{l_1}p_2^{l_2}p_3^{l_3}e^{i\nu f} 
\end{equation*}
to simplify the form of the generating functions of the Lie series transformations discussed later on, and so the above series is rewritten using the following trigonometric identity
\begin{equation*}
\cos(\nu f) = \frac{1}{2}e^{i \alpha f} + \frac{1}{2}e^{-i\alpha f},
\end{equation*}
yielding the following.
\begin{equation}
\begin{aligned}
  \frac 1{1+e\cos f} \approx \frac 1{\sqrt{1-e^2}}\left[e^0+\sum_{\nu=1}^{\frac 1 2 N_f-1}\eta^\nu e^{i \nu f}+\sum_{\nu=1}^{\frac 1 2 N_f-1}\eta^\nu e^{-i\nu f}\right]
\end{aligned}
\end{equation}
\begin{equation}
\frac{e\cos f}{1+e\cos f} \approx \frac{e}{\sqrt{1-e^2}}\left[\eta e^0 + \sum_{\nu=1}^{\frac 1 2 N_f-1}\left(\eta^{\nu+1}+\eta^{\nu-1}\right)e^{i\nu f}+\sum_{\nu=1}^{\frac 1 2 N_f-1}\left(\eta^{\nu+1}+\eta^{\nu-1}\right)e^{-i\nu f}\right]
\end{equation}
Here, $N_f=2^k$ is a user-specified truncation parameter.

\subsection{Floquet Transformation}
\label{subsec2}
The goal is to find the time-varying, periodic change of variables $P(f)$ that removes the time dependence from the quadratic part of the Hamiltonian.
Let the full state vector be denoted by $\mathbf{z} = [\mathbf{q},\mathbf{p}]^T$. The linear time-varying system corresponding to the quadratic part of the Hamiltonian can be written as
\begin{equation*}
\mathbf{z}' =A(f)\mathbf{z},
\end{equation*}
and the old state vector $\mathbf{z}$ can be expressed as a function of some new variables $\hat{\mathbf z} = [\hat{\mathbf q},\hat{\mathbf p}]^T$ via $P(f)$.
\begin{equation*}
\mathbf{z} = P(f)\hat{\mathbf{z}}
\end{equation*}
Differentiating the above expression with respect to $f$ yields the following,
\begin{equation*}
\mathbf{z}' = P'(f)\hat{\mathbf{z}} + P(f)\hat{\mathbf{z}}'
\end{equation*}
which can then be rewritten in terms of the new variables.
\begin{equation*}
\hat{\mathbf{z}}' = P(f)^{-1}\left(A(f)P(f) - P'(f)\right) \hat{\mathbf{z}}
\end{equation*}
To determine the derivative of $P(f)$ with respect to $f$ when $P(f)$ itself is not known, we must find an expression for $P(f)$ in terms of other quantities.
Let the state transition matrix (STM) of the LTV system at any true anomaly be denoted $\Phi(f_i)$. This matrix can be found by taking $\Phi(0) = I_{6\times 6}$ and integrating the following equation from $f=0$ to $f=f_i$.
\begin{equation*}
\Phi'(f) = A(f)\Phi(f)
\end{equation*}
From Floquet theory, we have the following relation involving the STM, a constant matrix $B$, and the desired transformation $P(f)$.
\begin{equation}\label{eq:floquet}
\Phi(f) = P(f)e^{fB}
\end{equation}
The matrix $B$ can be found by taking the logarithm of the monodromy matrix (i.e., $\Phi(2\pi)$). 
The above equation is equivalent to
\begin{equation*}
P(f) = \Phi(f)e^{-fB},
\end{equation*}
which we can now differentiate to find $\dot{P}(f)$.
\begin{equation*}
\begin{aligned}
P'(f) &= \Phi'(f)e^{-fB}-\Phi(f)e^{-fB}B\\
&= A(f)\Phi(f)e^{-fB}-\Phi(f)e^{-fB}B \\
&= A(f)P(f)-P(f)B
\end{aligned}
\end{equation*}
Plugging this expression for $\dot{P}(f)$ into the earlier equation yields
\begin{align*}
\hat{\mathbf{z}}' &= P(f)^{-1}\left(A(f)P(f) -A(f)P(f)+P(f)B\right)\hat{\mathbf{z}} \\
&= B\hat{\mathbf{z}}
\end{align*}
which results in the following LTI system.
\begin{equation*}
\hat{\mathbf{z}}' = B\hat{\mathbf{z}}
\end{equation*}
The rest of this section details the steps required to obtain the change of variables $P(f)$ for the ER3BP at either $L_1$ or $L_2$.
\subsubsection{Defining the Quadratic Part of the Hamiltonian}
To increase accuracy, the explicit form of the Floquet transformation should be found without performing the Fourier expansions. The quadratic part of the Hamiltonian after performing the Taylor expansion is of the form
\begin{equation}
H_2 = \frac{1}{2}\left(p_1^2+p_2^2+p_3^2\right) - p_2q_1+p_1q_2 + \frac{\beta(-2q_1^2+q_2^2+q_3^2)}{1+e\cos f}+\frac{e\cos f(q_1^2+q_2^2+q_3^2)}{2(1+e\cos f)},
\end{equation}
where $\beta$ is 
\begin{equation}
\beta = \frac{1}{2}\left(\frac{1-\mu}{|x_L-\mu|^3}+\frac{\mu}{|1-\mu-x_L|^3}\right),
\end{equation}
and $x_L$ denotes the $x$-coordinate of the libration point
\begin{equation}
x_L = 1-\mu\mp\gamma,
\end{equation}
where the upper sign corresponds to $L_1$ and the lower to $L_2$.
Hamilton's canonical equations are the following
\begin{equation}
\begin{aligned}
q'_1 &= p_1+q_2\\
q'_2 &= p_2-q_1\\
q'_3&= p_3
\end{aligned}\qquad
\begin{aligned}
p'_1 &= p_2 + \frac{4\beta-e\cos f}{1+e\cos f}q_1\\
p'_2 &= -p_1 - \frac{2\beta+e\cos f}{1+e\cos f}q_2\\
p'_3 &= -\frac{2\beta +e\cos f}{1+e\cos f}q_3,
\end{aligned}
\end{equation}
which can be written in matrix form as $\dot{\mathbf{z}}=A(f)\mathbf{z}$ where $\mathbf{z}=[\mathbf q,\mathbf p]^T$ and 
\begin{equation}
A(f) = \begin{bmatrix}0&1&0&1&0&0\\ -1&0&0&0&1&0\\ 0&0&0&0&0&1 \\ \frac{4\beta-e\cos f}{1+e\cos f}&0&0&0&1&0\\ 0& -\frac{2\beta+e\cos f}{1+e\cos f}&0&-1&0&0 \\ 0&0&-\frac{2\beta+e\cos f}{1+e\cos f}&0&0&0 \end{bmatrix}.
\end{equation}
\subsubsection{Propagate and Store the State Transition Matrix}
The 
Propagate and store the STM at evenly-spaced values of true anomaly: $f_i \in [0,2\pi], i=0,1,2,...,2^{M}$
\begin{equation}
\Phi'(f) = A(f)\Phi(f),\quad \Phi(0) = I_{6\times6}
\end{equation}
This can be done by propagating the columns of $\Phi$ separately. The system is quite unstable, meaning that high precision routines are needed.
\subsubsection{Find the Logarithm of the Monodromy Matrix}
The monodromy matrix is merely the STM evaluated at $f=2\pi$.
\begin{equation}
\Phi_e = \Phi(2\pi)
\end{equation}
The logarithm of the monodromy matrix, $B$, satisfies the following equation.
\begin{equation}
\Phi_e = e^{2\pi B}
\end{equation}
To begin, the monodromy matrix is diagonalized with the matrix $S$, the columns of which are the eigenvectors of the monodromy matrix. The result is the following,
\begin{equation}
D_M = S^{-1}\Phi_eS = \begin{bmatrix} e^{2\pi\lambda}&0&0&0&0&0\\ 0&a_2+ib_2&0&0&0&0\\ 0&0&a_3+ib_3&0&0&0\\0&0&0&e^{-2\pi\lambda}&0&0 \\ 0&0&0&0&a_2-ib_2&0 \\0&0&0&0&0&a_3-ib_3 \end{bmatrix}
\end{equation}
where $|a_j\pm b_j|=1$. As a result of this equality, the complex eigenvalues can be expressed as an exponential
\begin{equation}
\begin{aligned}
a_2+ib_2 &= e^{i2\pi\Omega_2} = \cos(2\pi\Omega_2)+i\sin(2\pi\Omega_2)\\
a_3+ib_3 &= e^{i2\pi\Omega_3} = \cos(2\pi\Omega_3)+i\sin(2\pi\Omega_3)\\
\end{aligned}
\end{equation}
where the values $\Omega_2$ and $\Omega_3$ can be calculated in a straightforward manner.
\begin{equation}
\begin{aligned}
\Omega_2 &= \frac{1}{2\pi}\cos^{-1}(a_2) \\
\Omega_3 &= \frac{1}{2\pi}\cos^{-1}(a_3)
\end{aligned}
\end{equation}
Due to the use of the logarithm, the values $\Omega_2$ and $\Omega_3$ are not unique ($e^{i2\pi\Omega} = e^{i2\pi(\Omega+k)},\,k\in \mathbb{Z})$). We could choose any  $k_2,k_3\in\mathbb{Z}$ to add to these frequencies and still recover the monodromy matrix by inverting the preceding steps. In practice, integers $k_2,k_3$ are chosen such that the transformation $P(f)$ is as close to identity as possible. For the Earth-Moon $L_1$ libration point, for example, values of $k_2=k_3=2$ are used.
The diagonal matrix consisting of the eigenvalues of the matrix $B$ that we seek can be written as 
\begin{equation}
D_B = \begin{bmatrix}\lambda&0&0&0&0&0\\ 0 & i(\Omega_2+k_2)& 0&0&0&0\\ 0 &0&i(\Omega_3+k_3)&0&0&0\\ 0&0&0&-\lambda&0&0\\
0&0&0&0&-i(\Omega_2+k_2)&0\\
0&0&0&0&0&-i(\Omega_3+k_3)\end{bmatrix},
\end{equation}
and the constant matrix $B$ can be constructed with $S$ and $D_B$.
\begin{equation}
B = SD_BS^{-1}
\end{equation}
To simplify future notation a bit, let $\omega_j =\Omega_j+k_j$ be the frequencies of the center subspaces.
One can ensure that the above steps have been followed correctly by checking that \eqref{eq:check} holds.
\begin{equation}\label{eq:check}
\Phi_e\cdot e^{-2\pi B} = I_{6\times 6}
\end{equation}
\subsubsection{Store Samples of the Floquet Transformation}
Through algebraic manipulation of \eqref{eq:floquet}, one obtains the following.
\begin{equation}
  P(f) = \Phi (f)e^{-fB}
\end{equation}
The STM is known at the sampled values of true anomaly, $f_i$, and the matrix $B$ is constant, so the Floquet transformation $P(f)$ can be computed and stored at the same evenly spaced $f_i$.
\begin{equation}
  P(f_j) = \Phi(f_j)e^{-f_jB}, \quad j=0,1,\dots,N_f-1
\end{equation}
\subsubsection{Fourier Approximation of the Floquet Transformation}

With the computed values of $P(f_j)$, all 36 elements of $P(f)$ can be approximated with Fourier series.
\begin{equation}
\mathcal{P}_{j,k} = \frac{1}{N_f}\sum^{N_f-1}_{n=0}p_{j,k}(f_n)e^{-if_n}
\end{equation}
\begin{equation}
p_{j,k}(f_n) = \sum^{N_f-1}_{n=0}\mathcal{P}_{j,k}e^{if_n}
\end{equation}
Here, $p_{j,k}(f_n)$ is the element in the $j$th row and $k$th column of the matrix $P(f_n)$.
\subsubsection{Applying the Floquet Transformation}
The symplectic Floquet transformation $P(f)$ cannot be directly substituted into the quadratic part of the Hamiltonian because it depends on $f$.
Instead, the symplectic Floquet transformation described in \cite{gomez2001dynamics} is applied.
To begin, the Hamiltonian is made autonomous by extending the phase space through the introduction of $F$, the conjugate momentum of $f$, to the Hamiltonian. The Hamiltonian is then in the following form.
\begin{equation}
H = F+H_2+H_3+\dots
\end{equation}
If the constant matrix $B$ is partitioned as
\begin{equation}
B = \begin{bmatrix}B_{11}&B_{12}\\B_{21}&B_{22}\end{bmatrix},
\end{equation}
then \cite{gomez2001dynamics} proves that there exists a symplectic transformation that brings the Hamiltonian to the form
\begin{equation}
\hat{H} = \hat{F} - \frac{1}{2}\hat{\mathbf{q}}^TB_{21}\hat{\mathbf{q}}-\hat{\mathbf{q}}^TB_{22}\hat{\mathbf{p}}+\frac{1}{2}\hat{\mathbf{p}}^TB_{12}\hat{\mathbf{p}} + \hat{H}_3+\hat{H}_4+\dots
\end{equation}
where the higher degree terms can be found by simply plugging in the Floquet change of variables. 
\begin{equation*}
\begin{bmatrix}\mathbf{q}\\\mathbf{p}\end{bmatrix} = P(f)\begin{bmatrix}\hat{\mathbf{q}}\\\hat{\mathbf{p}}\end{bmatrix}
\end{equation*}
The result of the theorem can be applied without explicitly finding the full symplectic transformation.

\subsection{Diagonalization}
\label{subsec3}

Once the quadratic part of the Hamiltonian has been made autonomous, it can be diagonalized in preparation for the Birkhoff transformations later on.
We seek a symplectic matrix $D$ that diagonalizes the quadratic Hamiltonian through the change of variables $\mathbf z = D\hat{\mathbf z}$. If the columns of $S$ are denoted as
\begin{equation}
S = \left[\mathbf{u}_1,\mathbf{u}_2+i\mathbf{v}_2,\mathbf{u}_3+i\mathbf{v}_3, \mathbf{v}_1,\mathbf{u}_2-i\mathbf{v}_2,\mathbf{u}_3-i\mathbf{v}_3 \right],
\end{equation}
then the symplectic matrix $D$ will be
\begin{equation}
D = \left[\kappa_1\mathbf{u}_1,\kappa_2\mathbf{u}_2,\kappa_3\mathbf{u}_3,\delta_1\kappa_1 \mathbf{v}_1,\delta_2\kappa_2\mathbf{v}_2,\delta_3\kappa_3\mathbf{v}_3 \right]
\end{equation}
where $\kappa_j = \sqrt{\mathbf{u}_j^TJ\mathbf{v}_j}$ and $\delta_j = \textrm{sgn}\left\{\mathbf{u}_j^TJ\mathbf{v}_j\right\}$, with $J$ being the standard $6\times 6$ symplectic matrix \cite{cabral2023normal}.
\begin{equation}
J = \begin{bmatrix}0& I_{3\times 3}\\-I_{3\times 3}&0\end{bmatrix}
\end{equation}
The linear symplectic matrix $D$ brings the quadratic part of the Hamiltonian to its real normal form.
\begin{equation}
\begin{bmatrix}\mathbf{q}\\\mathbf{p}\end{bmatrix} = D\begin{bmatrix}\hat{\mathbf{q}}\\\hat{\mathbf{p}}\end{bmatrix}
\end{equation}
\begin{equation}
\hat{H} = \hat{F}+\lambda \hat{q}_1\hat{p}_1 + \frac{\omega_2}{2}\left(\hat{q}_2^2+\hat{p}_2^2\right)+ \frac{\omega_3}{2}\left(\hat{q}_3^2+\hat{p}_3^2\right)+\hat{H}_3+\hat{H}_4+\dots
\end{equation}
\subsection{Complexification}
\label{subsec4}

The following transformation is applied to cast the Hamiltonian into complexified coordinates
\begin{equation}
\begin{bmatrix}q_1\\ p_1\end{bmatrix} = I_{2\times 2}\begin{bmatrix}\hat{q}_1\\\hat{p}_1\end{bmatrix} \quad
\begin{bmatrix}q_2\\ p_2\end{bmatrix} = \frac{1}{\sqrt{2}}\begin{bmatrix}1&i\\i&1\end{bmatrix}\begin{bmatrix}\hat{q}_2\\\hat{p}_2\end{bmatrix} \quad
\begin{bmatrix}q_3\\ p_3\end{bmatrix} = \frac{1}{\sqrt{2}}\begin{bmatrix}1&i\\i&1\end{bmatrix}\begin{bmatrix}\hat{q}_3\\\hat{p}_3\end{bmatrix}
\end{equation}
resulting in a new Hamiltonian of the following form.
\begin{equation}
\hat{H} = \hat{F} + \lambda \hat{q}_1\hat{p}_1 + i\omega_2\hat{q}_2\hat{p}_2 + i\omega_3\hat{q}_3\hat{p}_3 + \hat{H}_3+\hat{H}_4+\dots
\end{equation}
The benefit of this transformation is only obvious in retrospect. The generating functions for the Lie series transformations that are about to be discussed are much easier to find when the quadratic part of the Hamiltonian is in the above form. For now, it suffices to know that the complexification results in a diagonal linear dynamical system for the quadratic part of the Hamiltonian
\begin{equation}
\begin{bmatrix} q_1'\\ q_2'\\ q_3'\\ p_1'\\ p_2'\\ p_3'\end{bmatrix} = \begin{bmatrix}\lambda&0&0&0&0&0\\0&i\omega_2&0&0&0&0\\0&0&i\omega_3&0&0&0\\
0&0&0&-\lambda&0&0\\0&0&0&0&-i\omega_2&0\\0&0&0&0&0&-i\omega_3\end{bmatrix}\begin{bmatrix}q_1\\q_2\\q_3\\p_1\\p_2\\p_3\end{bmatrix}
\end{equation}
which follows directly from Hamilton's equations.

\subsection{Lie Series Transformations}
\label{subsec5}
In the extended phase space, the Poisson bracket of a Hamiltonian $H$ and a generating function $G$ is defined as
\begin{equation}
\{H,G\} = \left[\sum_{j=1}^3\frac{\partial H}{\partial q_j}\frac{\partial G}{\partial p_j} - \frac{\partial H}{\partial p_j}\frac{\partial G}{\partial q_j}\right] + \frac{\partial H}{\partial f}\frac{\partial G}{\partial F} - \frac{\partial H}{\partial F}\frac{\partial G}{\partial f}.
\end{equation}
If $G$ does not depend explicitly on the conjugate momentum  $F$, then the Poisson bracket becomes the following.
\begin{equation}
\{H,G\} = \left[\sum_{j=1}^3\frac{\partial H}{\partial q_j}\frac{\partial G}{\partial p_j} - \frac{\partial H}{\partial p_j}\frac{\partial G}{\partial q_j}\right]  - \frac{\partial H}{\partial F}\frac{\partial G}{\partial f}
\end{equation}
Consider the terms of the Hamiltonian up to and including degree two. Since the conjugate momenta $F$ only appears on its own, the Poisson bracket of the terms up to degree two and a generating function will be
\begin{equation}
\{F+H_2,G\} = \{H_2,G\}-\frac{\partial G}{\partial f}.
\end{equation}
A generating function can be applied to a Hamiltonian through a Lie series transformation, denoted $H\circ \Phi^G$, yielding a new Hamiltonian.
\begin{equation}
\begin{aligned}
\hat H&= H+\left\{H,G\right\}+\frac 1 2\left\{\left\{H,G\right\},G\right\}+\frac{1}{3!}\{\{\{H,G\},G\},G\}\dots\\
&= H \circ\Phi^{G}
\end{aligned}
\end{equation}
The goal is to remove certain terms from the Hamiltonian up to the truncation degree N. This will be accomplished through the Dragt-Finn method, which allows one to define a sequence of $N-2$ Lie series transformations that remove terms of increasing degree from the Hamiltonian. For example, the generating function $G_3$ will remove terms of order 3, $G_4$ terms of degree 4, and so on.
Applying the first generating function, $G_3$, yields the new Hamiltonian $H^{(3)}$\cite{jorba1998numerical}.
\begin{equation}
\begin{aligned}
H^{(3)} &= H\circ\Phi^{G_3}\\
 &= F^{(3)} + H_2^{(3)}+H_3^{(3)}+H_4^{(3)}+\dots
\end{aligned}
\end{equation}
Using the above definition of the Poisson bracket in the extended phase space, the terms of each degree within the new Hamiltonian can be collected as follows.
\begin{equation}
\begin{aligned}
F^{(3)} &= F\\
H_2^{(3)} &= H_2 \\
H_3^{(3)} &= H_3+\{F+H_2,G_3\}\\
H_4^{(3)} &= H_4+\{H_3,G_3\}+\frac{1}{2!}\{\{F+H_2,G_3\},G_3\}\\
H_5^{(3)} &= H_5 + \{H_4,G_3\}+\frac{1}{2!}\{\{H_3,G_3\},G_3\}+\frac{1}{3!}\{\{\{F+H_2,G_3\},G_3\},G_3\}\\
\vdots
\end{aligned}
\end{equation}
Note that the conjugate momentum $F$ as well as the quadratic part of the Hamiltonian remain unchanged by this transformation.
It turns out that all terms of degree 3 can be removed from the Hamiltonian in this manner. Next, $G_4$ will be applied, yielding
\begin{equation}
\begin{aligned}
H^{(4)} &= H^{(3)}\circ\Phi^{G_4}\\
&= H\circ\Phi^{G_3}\circ\Phi^{G_4}\\
 &= F^{(4)} + H_2^{(4)}+H_3^{(4)}+H_4^{(4)}+\dots
\end{aligned}
\end{equation}
where
\begin{equation}
\begin{aligned}
F^{(4)} &= F^{(3)}\\
H_2^{(4)} &= H^{(3)}_2 \\
H_3^{(4)} &= H^{(3)}_3= 0\\
H_4^{(4)} &= H^{(3)}_4+\{F+H^{(3)}_2,G_4\}\\
H_5^{(4)} &= H^{(3)}_5 \\
H_6^{(4)} &= H^{(3)}_6 + \{H^{(3)}_4,G_4\}+\frac{1}{2!}\{\{F+H_2^{(3)},G_4\},G_4\}\\
\vdots
\end{aligned}
\end{equation}
This process is continued up to degree N, leaving the Hamiltonian in the form
\begin{equation}
H^{(N)} = H_2^{(N)}(q_1p_1,q_2p_2,q_3p_3)+H_4^{(N)}(q_1p_1,q_2p_2,q_3p_3)+H^{(N)}_6(q_1p_1,q_2p_2,q_3p_3)+\dots
\end{equation}
up to degree $N$ or $N-1$ depending on whether $N$ is even or odd.
The next subsection discusses the form of the generating functions and addresses which terms can and cannot be removed from the Hamiltonian through Lie series transformations.
\subsubsection{Defining Generating Functions}
Let the $k$th degree generating function be defined as the collection of monomials
\begin{equation}
G_k = \sum_{(\mathbf{m},\mathbf{l},\nu)\in \mathcal{K}}g^{\mathbf{m},\mathbf{l}}_\nu q_1^{m_1}q_2^{m_2}q_3^{m_3}p_1^{l_1}p_2^{l_2}p_3^{l_3}e^{i\nu f},
\end{equation}
where
\begin{equation}
\mathcal{K} = \left\{\mathbf{m},\mathbf l\in \mathbb{N}^3, \nu \in \mathbb{Z}\,\left|\, \sum_{j=1}^3(m_j+l_j)=k \quad \textrm{and}\quad |\nu| < \frac{N_f} 2-1\right.\right\}.
\end{equation}
Through the bilinearity of the Poisson bracket, the following holds.
\begin{equation}
\left\{F+H_2,G_k\right\} =\sum_{(\mathbf{m},\mathbf{l},\nu)\in \mathcal{K}} \left\{F+H_2,g^{\mathbf{m},\mathbf{l}}_\nu q_1^{m_1}q_2^{m_2}q_3^{m_3}p_1^{l_1}p_2^{l_2}p_3^{l_3}e^{i\nu f}\right\}
\end{equation}
Let us evaluate this Poisson bracket for any particular element of the generating function.
\begin{equation}
\begin{aligned}
\left\{F+H_2,g^{\mathbf{m},\mathbf{l}}_\nu q_1^{m_1}q_2^{m_2}q_3^{m_3}p_1^{l_1}p_2^{l_2}p_3^{l_3}e^{i\nu f}\right\}
\end{aligned}
\end{equation}
This can be broken up into the following two terms
\begin{equation}
 \left\{F,g^{\mathbf{m},\mathbf{l}}_\nu q_1^{m_1}q_2^{m_2}q_3^{m_3}p_1^{l_1}p_2^{l_2}p_3^{l_3}e^{i\nu f}\right\}= -i\nu g^{\mathbf{m},\mathbf{l}}_\nu q_1^{m_1}q_2^{m_2}q_3^{m_3}p_1^{l_1}p_2^{l_2}p_3^{l_3}e^{i\nu f}
\end{equation}
\begin{equation}
\left\{H_2,g^{\mathbf{m},\mathbf{l}}_\nu q_1^{m_1}q_2^{m_2}q_3^{m_3}p_1^{l_1}p_2^{l_2}p_3^{l_3}e^{i\nu f}\right\} =  (\mathbf{l}-\mathbf{m})^T\boldsymbol{\eta}\cdot g^{\mathbf{m},\mathbf{l}}_\nu q_1^{m_1}q_2^{m_2}q_3^{m_3}p_1^{l_1}p_2^{l_2}p_3^{l_3}e^{i\nu f},
\end{equation}
where $\boldsymbol{\eta}=[\lambda,i\omega_2,i\omega_3]^T$.
Note that the exponents and frequency are unchanged. This is an extremely useful result, as it means that the term of the generating function with coefficient $g^{\mathbf{m},\mathbf{l}}_\nu$ depends only on the term of the Hamiltonian with coefficient $h^{\mathbf{m},\mathbf{l}}_\nu$. This is the motivation for the earlier complexification transformation and follows directly from the diagonal form of the dynamical linear system.
In the case where $k=3$, we would like to remove all of the third-degree terms from the Hamiltonian. In other words, we would like for $H_3^{(3)}=0$
\begin{equation}
\begin{aligned}
H_3^{(3)} &= H_3 + \{F+H_2,G_3\}\\
0 &= H_3+\{F+H_2,G_3\}
\end{aligned}
\end{equation}
Grouping coefficients with the same exponents and frequencies yields
\begin{equation}
h^{\mathbf{m},\mathbf{l}}_\nu + \left((\mathbf{l}-\mathbf{m})^T\boldsymbol{\eta}-i\nu\right)g^{\mathbf{m},\mathbf{l}}_\nu=0,
\end{equation}
which can be manipulated to obtain an expression for the coefficients of the generating function in terms of the coefficients of the Hamiltonian.
\begin{equation}\label{eq:genfunc}
g^{\mathbf{m},\mathbf{l}}_\nu =\frac{-h^{\mathbf{m},\mathbf{l}}_\nu}{(\mathbf{l}-\mathbf{m})^T\boldsymbol{\eta}-i\nu}
\end{equation}
This same methodology can be used without issue for odd-degree terms, but for some terms of even degree, the denominator will be equal to zero. This occurs when (assuming there is no resonance relation between the elements of $\boldsymbol{\eta}$  up to order $N$) the exponents and frequency of a monomial satisfy the following equalities.
\begin{equation}
\mathbf l - \mathbf m = 0\quad \textrm{and}\quad \nu = 0
\end{equation}
The terms that satisfy this resonance condition cannot be removed from the Hamiltonian, and thus must be kept.
This process of removing non-resonant terms is repeated up to some finite degree, in this case $N$.
The result is a Hamiltonian of the form
\begin{equation}
H^{(N)}= H^{(N)}_2(q_1p_1,q_2p_2,q_3p_3)+H^{(N)}_4(q_1p_1,q_2p_2,q_3p_3)+H^{(N)}_6(q_1p_1,q_2p_2,q_3p_3)+\dots
\end{equation}

\subsection{Realification}
\label{subsec6}
The Hamiltonian can be transformed from its complex normal form to its real normal form by applying the inverse of the complexifying transformation from earlier. The normal form state is denoted $\mathbf x_{NF} = [\tilde x,\tilde y,\tilde z,\tilde p_x,\tilde p_y,\tilde p_z]^T$
\begin{equation}
\begin{bmatrix}q_1\\ p_1\end{bmatrix} = I_{2\times 2}\begin{bmatrix}\tilde x\\\tilde p_x\end{bmatrix} \quad
\begin{bmatrix}q_2\\ p_2\end{bmatrix} = \frac{1}{\sqrt{2}}\begin{bmatrix}1&-i\\-i&1\end{bmatrix}\begin{bmatrix}\tilde y\\\tilde p_y\end{bmatrix} \quad
\begin{bmatrix}q_3\\ p_3\end{bmatrix} = \frac{1}{\sqrt{2}}\begin{bmatrix}1&-i\\-i&1\end{bmatrix}\begin{bmatrix}\tilde z\\\tilde p_z\end{bmatrix}
\end{equation}
The result is a Hamiltonian of the form
\begin{equation}
H^{(N)} = H_2^{(N)}\left(\tilde x\tilde p_x, \frac{1}{2}\left( \tilde y^2 + \tilde p_y^2 \right), \frac 1 2 \left(\tilde z^2+\tilde p_z^2\right)\right) + H_4^{(N)}\left(\tilde x\tilde p_x, \frac{1}{2}\left( \tilde y^2 + \tilde p_y^2 \right), \frac 1 2 \left(\tilde z^2+\tilde p_z^2\right)\right)+\dots,
\end{equation}
The generating functions $G_3,G_4,\dots,G_N$ can also be realified at this point. This is done so that the coordinate transformations that will be defined shortly can be performed in the realified coordinates.

\subsection{Action-Angle Variables}
\label{subsec7}
The following canonical transformation, $\textrm f_{AA}^{-1}: \mathbf{x}_{NF}\to\mathbf{x}_{AA}$, results in a Hamiltonian that is only a function of the actions denoted $I_k,\,k\in1,2,3$.
\begin{equation}
\begin{aligned}
I_1 &= \left|\tilde{x}\tilde{p}\right|\\
I_2 &= \frac{1}{2}\left(\tilde y^2+\tilde p_y^2\right)\\
I_3 &= \frac 1 2\left(\tilde z^2 + \tilde p_z^2\right)
\end{aligned}\qquad
\begin{aligned}\label{eq:fAA}
\phi_1 &= \frac{1}{2}\ln \left(\left|\frac{\tilde x}{\tilde p_x}\right|\right)\\
\phi_2 &= -\tan^{-1}\left(\frac{\tilde p_y}{\tilde y}\right)\\
\phi_2 &= -\tan^{-1}\left(\frac{\tilde p_z}{\tilde z}\right)
\end{aligned}
\end{equation}
\begin{equation}
H_{AA} = H_1(I_1,I_2,I_3)+H_2(I_1,I_2,I_3)+\dots+H_N(I_1,I_2,I_3)
\end{equation}
\begin{equation}
H_n(I_1,I_2,I_3) = \sum_{|\mathbf k|_1=n}h_{\mathbf k}I_1^{k_1}I_2^{k_2}I_3^{k_3}, \quad \mathbf k \in \mathbb N^3
\end{equation}
Note that the degree of every term is reduced by a factor of two as a result of the action-angle transformation. If the truncation degree is $N=10$, for example, the final action-angle Hamiltonian will have terms up to degree five.
The angles $\phi_k,\,k\in1,2,3$ are therefore linearly varying with true anomaly $f$ while the actions are constant.
\begin{equation}
\begin{aligned}
  \frac{\textrm d \phi_1}{\textrm d f} &= \frac{\partial H_{AA}}{\partial I_1} = c_1\\
  \frac{\textrm d \phi_2}{\textrm d f} &= \frac{\partial H_{AA}}{\partial I_2} = c_2\\
  \frac{\textrm d \phi_3}{\textrm d f} &= \frac{\partial H_{AA}}{\partial I_3} = c_3
\end{aligned}\qquad
\begin{aligned}
  \frac{\textrm d I_1}{\textrm d f} &= -\frac{\partial H_{AA}}{\partial \phi_1} = 0\\
  \frac{\textrm d I_2}{\textrm d f} &= -\frac{\partial H_{AA}}{\partial \phi_2} = 0\\
  \frac{\textrm d I_3}{\textrm d f} &= -\frac{\partial H_{AA}}{\partial \phi_3} = 0
\end{aligned}
\end{equation}
\begin{equation}
\begin{aligned}
\phi_1(f) &= c_1f+\phi_1(0)\\
\phi_2(f) &= c_2f+\phi_2(0)\\
\phi_3(f) &= c_3f+\phi_3(0)
\end{aligned}\qquad
\begin{aligned}
I_1(f)&=I_1(0)\\
I_2(f)&=I_2(0)\\
I_3(f)&=I_3(0)\\
\end{aligned}
\end{equation}
The inverse action angle transformation $\textrm f_{AA}:\mathbf x_{AA}\to\mathbf x_{NF}$ is included below for completeness.
\begin{equation}
\begin{aligned}
\tilde x &= \sqrt{I_1}e^{\phi_1}\\
\tilde y &= \sqrt{2I_2}\cos(\phi_2)\\
\tilde z &= \sqrt{2I_3}\cos(\phi_3)
\end{aligned} \qquad
\begin{aligned}
\tilde p_x &= \sqrt{I_1}e^{-\phi_1}\\
\tilde p_y &= -\sqrt{2I_2}\sin(\phi_2)\\
\tilde p_z &= -\sqrt{2I_3}\sin(\phi_3)
\end{aligned}
\end{equation}

\subsection{Coordinate Transformations}
To begin, we denote the original state vector in the rotating-pulsating frame as $\mathbf{x}_{RTB} = [x,y,z,x',y',z']^T$. We then apply an affine transformation to place the desired libration point at the origin of the new coordinate frame, followed by a linear transformation $V$ that brings the state to the canonical coordinates and conjugate momenta about the libration point, denoted $\mathbf x_\ell = [q_1,q_2,q_3,p_1,p_2,p_3]^T$ . This can be summarized as the following affine transformation
\begin{equation}
\mathbf{x}_{RTB} = V(T\mathbf{x}_\ell+\mathbf{b})
\end{equation}
where
\begin{equation}
  V = \left[\begin{matrix}1&0&0&0&0&0\\0&1&0&0&0&0\\0&0&1&0&0&0\\0&1&0&1&0&0\\-1&0&0&0&1&0\\0&0&0&0&0&1\end{matrix}\right],\quad T=\begin{bmatrix}\delta\gamma&0&0&0&0&0\\0&\delta\gamma&0&0&0&0\\0&0&\gamma&0&0&0\\0&0&0&\delta\gamma&0&0\\0&0&0&0&\delta\gamma&0\\0&0&0&0&0&\gamma\end{bmatrix},\quad\text{and}\quad \mathbf{b} = \begin{bmatrix}1-\mu\\0\\0\\0\\a-\mu\\0\end{bmatrix}.
\end{equation}
Here, $\gamma$ is the distance from the libration point to the nearest primary and $\mu$ is the mass parameter of the system. For $L_1$ and $L_2$, $a=1\mp\gamma$ and $\delta = 1$, where the upper sign corresponds to $L_1$ and the lower to $L_2$. For $L_3$, $a=-\gamma$ and $\delta=-1$.
After transforming to the scaled and translated Hamiltonian state, the Floquet transformation, $P(f)$, is applied along with the diagonalizing transformation, $D$, yielding a real diagonalized state that can be denoted $\mathbf x_{qp}\equiv \mathbf x_{qp}^{(2)} = \left[q_1^{(2)},q_2^{(2)},q_3^{(2)},p_1^{(2)},p_2^{(2)},p_3^{(2)}\right]^T$. 
\begin{equation}
\mathbf{x}_{\ell} = P(f)D\mathbf x_{qp}^{(2)}
\end{equation}
The superscript does not mean anything yet, but it will be used later on to keep track of the number of generating functions $G_k,\,k\in3,\dots,N$ that have been applied to the coordinates.
The full transformation between the real diagonalized state and the original rotating-pulsating state is denoted as $\textrm g:\mathbf x_{qp}^{(2)}\to \mathbf x_{RTB}$ and is shown in \eqref{eq:g}. The inverse transformation $\textrm g^{-1}: \mathbf x_{RTB}\to\mathbf x_{qp}^{(2)}$ is then \eqref{eq:ginv}.
\begin{equation}\label{eq:g}
\mathbf{x}_{RTB}=\textrm g\left(\mathbf{x}_{qp}^{(2)}\right) = V\left(TP(f)D\mathbf{x}_{qp}^{(2)}+\mathbf b\right) 
\end{equation}
\begin{equation}\label{eq:ginv}
  \mathbf{x}_{qp}^{(2)} = \textrm g^{-1}\left(\mathbf x_{RTB}\right)= D^{-1}P(f)^{-1}T^{-1}\left(V^{-1}\mathbf{x}_{RTB} - \mathbf{b}\right) 
\end{equation}
Next, the Lie series transformations are applied in sequence with the \textit{realified} generating functions.
\begin{equation}
\begin{aligned}
H^{(3)}&= H^{(2)}+\left\{H^{(2)},G_3\right\}+\frac 1 2\left\{\left\{H^{(2)},G_3\right\},G_3\right\}+\dots\\
&= H^{(2)} \circ\Phi^{G_3}
\end{aligned}
\end{equation}
\begin{equation}
H^{(N)} = H^{(2)}\circ\Phi^{G_3}\circ\Phi^{G_4}\circ\dots\circ\Phi^{G_N}
\end{equation}
We will denote the state immediately after the $k$th generating function has been applied to be $\mathbf x_{qp}^{(k)}$. After applying all of the generating functions, the result is $\mathbf x_{qp}^{(N)}\equiv\mathbf x_{NF}$, the normal form coordinates.
Let us define $\textrm{\textbf{q}}_j^{(n)}$ to be the function that maps from the state $\mathbf x_{qp}^{(n)}$ to the original coordinate $q_j^{(2)}$ (i.e., $\textrm{\textbf{q}}_j^{(n)}:\mathbf{x}_{qp}^{(n)} \to q_j^{(2)}$). Then, $\textrm {\textbf{q}}^{(2)}_j$ can simply be viewed as a function that selects the $j$th coordinate of the $\mathbf{x}_{qp}^{(2)}$ state vector. To obtain polynomial expressions for the original $q_j$ in terms of the normal form coordinates, one can apply the generating functions in order (i.e., $G_3,G_4,\dots,G_N$ ) to the function $\textrm {\textbf{q}}_j^{(2)}$.
\begin{equation}
\textrm {\textbf{q}}_j^{(N)} = \textrm {\textbf{q}}_j^{(2)}\circ\Phi^{G_3}\circ\Phi^{G_4}\circ\dots\circ\Phi^{G_N}(\mathbf x_{NF})
\end{equation}
In other words, the same sequence of Lie series transformations that was applied to the Hamiltonian can be applied to $\textrm {\textbf{q}}_j^{(2)}$, yielding the function $\textrm {\textbf{q}}_j^{(N)}:\mathbf x_{NF}\to q^{(2)}_j$. This same process can be applied to the functions $\textrm{\textbf{p}}_j^{(2)}$ that output the conjugate momenta.
\begin{equation}
\begin{aligned}
\textrm {\textbf{q}}^{(n)}_j &= \textrm {\textbf{q}}^{(n-1)}_j +\left\{\textrm {\textbf{q}}^{(n-1)}_j,G_n\right\}+\frac{1}{2!}\left\{\left\{\textrm {\textbf{q}}^{(n-1)}_j,G_n\right\},G_n\right\}+\dots,\quad j=1,2,3\\
\textrm {\textbf{p}}^{(n)}_j &= \textrm {\textbf{p}}^{(n-1)}_j +\left\{\textrm {\textbf{p}}^{(n-1)}_j,G_n\right\}+\frac{1}{2!}\left\{\left\{\textrm {\textbf{p}}^{(n-1)}_j,G_n\right\},G_n\right\}+\dots,\quad j=1,2,3
\end{aligned}
\end{equation}
Let the function that applies the sequence of Lie series transformations be denoted as
\begin{equation}
\mathcal{T} = \Phi^{G_3}\circ \Phi^{G_4}\circ\dots\circ\Phi^{G_N},
\end{equation}
then the inverse transformation will be the following
\begin{equation}
\mathcal{T}^{-1} = \Phi^{-G_N}\circ\dots\circ \Phi^{-G_4}\circ\Phi^{-G_3},
\end{equation}
which can be used to find polynomial expressions for the normal form coordinates in terms of the original Hamiltonian state. 
\begin{equation}
\begin{aligned}
\textrm {\textbf{q}}^{(n-1)}_j &= \textrm {\textbf{q}}^{(n)}_j +\left\{\textrm {\textbf{q}}^{(n)}_j,-G_{n}\right\}+\frac{1}{2!}\left\{\left\{\textrm {\textbf{q}}^{(n)}_j,-G_{n}\right\},-G_{n}\right\}+\dots,\quad j=1,2,3\\
\textrm {\textbf{p}}^{(n-1)}_j &= \textrm {\textbf{p}}^{(n)}_j +\left\{\textrm {\textbf{p}}^{(n)}_j,-G_n\right\}+\frac{1}{2!}\left\{\left\{\textrm {\textbf{p}}^{(n)}_j,-G_n\right\},-G_n\right\}+\dots,\quad j=1,2,3
\end{aligned}
\end{equation}
Here, the generating functions must be applied in the order $-G_N,\dots,-G_4,-G_3$.
A normal form state $\mathbf x_{NF}$ can be transformed to the original diagonal state by simply applying the operator $\mathcal{T}$
\begin{equation}
\mathbf{x}_{qp}^{(2)} = \mathcal{T}\left(\mathbf x_{NF}\right) = \Phi^{G_3}\circ \Phi^{G_4}\circ\dots\circ\Phi^{G_N}\left(\mathbf x_{NF}\right),
\end{equation}
and the inverse transformation can be accomplished with $\mathcal{T}^{-1}$.
\begin{equation}
\mathbf{x}_{NF} = \mathcal{T}^{-1}\left(\mathbf x_{qp}^{(2)}\right) = \Phi^{-G_N}\circ\dots\circ \Phi^{-G_4}\circ\Phi^{-G_3}\left(\mathbf x_{qp}^{(2)}\right)
\end{equation}
The full transformation that takes an action angle state and transforms it to a state in the rotating-pulsating restricted three-body state is denoted $\mathcal{A}$ (for ``analytical'') and can be written as the following.
\begin{equation}
\mathbf{x}_{RTB} = \mathcal{A}\left(\mathbf x_{AA}^B\right) = \textrm g\circ\mathcal{T}\circ \textrm f_{AA}\left(\mathbf x_{AA}^B\right)
\end{equation}
The inverse analytical transformation that takes a state in the rotating-pulsating frame and outputs an action-angle state is then
\begin{equation}
\mathbf{x}_{AA} = \mathcal{A}^{-1}(\mathbf x_{RTB}) = \textrm f_{AA}^{-1}\circ \mathcal{T}^{-1}\circ \textrm g^{-1}(\mathbf x_{RTB}).
\end{equation}
Unfortunately, the composition of the analytical transformation and its inverse, $\mathcal{T}^{-1}\circ\mathcal{T}$, poorly approximates the identity transformation.  This motivates a numerical approach to the transformation.
Rather than applying the sequence of Lie series transformations to obtain polynomial expressions for $\mathbf x_{qp}^{(2)}$ and $\mathbf x_{NF}$, one can instead numerically integrate states according to the equations of motion defined by the generating functions.
To perform the forward transformation, the generating functions are applied in the order $G_N,\dots,G_4,G_3$.
\begin{equation}
\mathbf x_{qp}^{(n-1)} = \mathbf x_{qp}^{(n)} + \int_0^{1}\left.\left\{\mathbf{x}_{qp},G_n\right\}\right|_{\mathbf{x}^{(n)}_{qp}(\tau)}d\tau
\end{equation}
The inverse of the numerical transformation is obtained by applying the generating functions in the order $-G_3,-G_4,\dots,-G_N$, which is equivalent to the following.
\begin{equation}
\mathbf x_{qp}^{(n)} = \mathbf x_{qp}^{(n-1)} + \int_0^{-1}\left.\left\{\mathbf{x}_{qp},G_n\right\}\right|_{\mathbf{x}^{(n-1)}_{qp}(\tau)}d\tau
\end{equation}
The resulting numerical forward transformations are not generally any better than the analytical transformation.
\begin{equation}
  \mathbf{x}_{RTB} = \mathcal{N}(\mathbf x_{AA}) = g\circ\mathcal{T}_{num}\circ \textrm f_{AA}\left(\mathbf x_{AA}^B\right)
\end{equation}
However, the numerical inverse transformation, denoted $\mathcal{N}^{-1}$, is more accurate because $\mathcal T^{-1}_{num}\circ \mathcal T_{num}$ is identity up to the integration tolerance specified.
\begin{equation}
\mathbf{x}_{AA} = \mathcal{N}^{-1}(\mathbf x_{RTB}) = \textrm f_{AA}^{-1}\circ \mathcal{T}^{-1}_{num}\circ \textrm g^{-1}(\mathbf x_{RTB})
\end{equation}
\section{The Resonant ER3BP Normal Form}
The Birkhoff normal form is convenient because it provides a simple dynamical model for the ER3BP. However, it generally suffers from a small region of validity due to the large number of terms that are removed from the Hamiltonian through Birkhoff Lie series transformations. This is especially apparent for trajectories out-of-plane motion. A consequence of this is that the Birkhoff normal form provides reasonable characterization of only a small section of the halo family, and no means at all of characterizing the quasihalo family of trajectories.
Let us return to the equation for the coefficients of the generating functions,
\begin{equation}
g^{\mathbf{m},\mathbf{l}}_\nu =\frac{-h^{\mathbf{m},\mathbf{l}}_\nu}{(\mathbf{l}-\mathbf{m})^T\boldsymbol{\eta}-i\nu}
\end{equation}
where $\boldsymbol{\eta}=[\lambda,i\omega_2,i\omega_3]^T$. If there was a first-order resonance between the center frequencies (i.e., $\omega_2=\omega_3$), then the terms of the Hamiltonian with exponents that satisfy the following condition would have to be kept.
\begin{equation}
l_1-m_1=0\quad\text{and}\quad(l_2-m_2)+(l_3-m_3)=0 \quad\text{and}\quad\nu=0
\end{equation}
It is clear from the numerical results included in Tables \ref{tab:L1Birkhoffcoeffs} and \ref{tab:resonantL2srp} that there is no such first-order resonance between the frequencies in the Earth-Moon or Sun-Earth (or even the Sun-EM barycenter) systems. However, the halo family of trajectories bifurcates from the planar Lyapunov family when the in-plane frequency matches the out-of-plane frequency, making this line of reasoning relevant.
Proceeding with the Lie series transformations while removing all terms except those that satisfy the above condition yields a Hamiltonian of the following form.
\begin{equation}
H_{AA} = H_1(I_1,I_2,I_3,\phi_2-\phi_3)+H_2(I_1,I_2,I_3,\phi_2-\phi_3)+H_3(I_1,I_2,I_3,\phi_2-\phi_3)+\dots
\end{equation}
From Hamilton's equations, this would result in varying $I_2$ and $I_3$ actions; however, an additional transformation can be applied to regain a constant of motion.
Let us define a new resonant action-angle state to be $\mathbf{x}_{AA}^R = \left[\hat I_1,\hat I_2,\hat I_3,\theta_1,\theta_2,\theta_3\right]^T$. The linear transformation taking the Birkhoff action-angle state to the new resonant action-angle state can be denoted $\textrm h^{-1}:\mathbf{x}_{AA}^B\to\mathbf x_{AA}^R$
\begin{equation}
\begin{aligned}
\hat{I}_1&=I_1\\
\hat{I}_2 &= I_2\\
\hat I_3 &= I_2+I_3
\end{aligned}\qquad
\begin{aligned}
\theta_1&=\phi_1\\
\theta_2 &= \phi_2-\phi_3\\
\theta_3 &= \phi_3
\end{aligned}
\end{equation}
and its inverse $\textrm h:\mathbf x_{AA}^R\to\mathbf x_{AA}^B$ is the following.
\begin{equation}
\begin{aligned}
I_1&=\hat I_1\\
I_2 &= \hat I_2\\
I_3 &= \hat I_3-\hat I_2
\end{aligned}\qquad
\begin{aligned}
\phi_1&=\theta_1\\
\phi_2 &= \theta_2+\theta_3\\
\phi_3 &= \theta_3
\end{aligned}
\end{equation}
Applying $\textrm h^{-1}$ to the Hamiltonian $H_{AA}$ yields a Hamiltonian that depends on only one of the angles.
\begin{equation}
H_{AA}^R = H_1\left(\hat I_1,\hat I_2,\hat I_3,\theta_2\right) + H_2\left(\hat I_1,\hat I_2,\hat I_3,\theta_2\right) + H_3\left(\hat I_1,\hat I_2,\hat I_3,\theta_2\right)+\dots
\end{equation}
As a result, the actions $\hat I_1$ and $\hat I_3$ will be constant. Since the $\hat I_2$ action will generally be fluctuating (except in some special cases), the angles will generally no longer be linearly varying.
\begin{equation}
\begin{aligned}
\frac{\textrm d \theta_1}{\textrm d f} &= \frac{\partial H_{AA}^R}{\partial \hat I_1} \neq \text{const.}\\
\frac{\textrm d \theta_2}{\textrm d f} &= \frac{\partial H_{AA}^R}{\partial \hat I_2} \neq \text{const.}\\
\frac{\textrm d \theta_3}{\textrm d f} &= \frac{\partial H_{AA}^R}{\partial \hat I_3} \neq \text{const.}
\end{aligned}\qquad
\begin{aligned}
\frac{\textrm d \hat I_1}{\textrm d f} &= -\frac{\partial H_{AA}^R}{\partial \theta_1} = 0\\
\frac{\textrm d \hat I_2}{\textrm d f} &= -\frac{\partial H_{AA}^R}{\partial \theta_2} \neq 0\\
\frac{\textrm d \hat I_3}{\textrm d f} &= -\frac{\partial H_{AA}^R}{\partial \theta_3} = 0
\end{aligned}
\end{equation}
We have essentially exchanged a constant of motion for a larger region of validity, and can now characterize a larger section of the halo family along with the family of quasihalo trajectories.
The full analytical and numerical transformations from the resonant action-angle state to the original ER3BP state in the rotating-pulsating frame can be summarized as
\begin{equation}
\begin{aligned}
  \mathbf{x}_{RTB} &= {}^R\mathcal{A}\left(\mathbf x_{AA}^R\right) = \textrm g\circ{}^R\mathcal{T}\circ \textrm f_{AA}\circ \textrm h\left(\mathbf x_{AA}^R\right)\\
  \mathbf{x}_{RTB} &= {}^R\mathcal{N}\left(\mathbf x_{AA}^R\right) = \textrm g\circ{}^R\mathcal{T}_{num}\circ \textrm f_{AA}\circ \textrm h\left(\mathbf x_{AA}^R\right)
\end{aligned}
\end{equation}
and the inverse transformations will be the following.
\begin{equation}
\begin{aligned}
  \mathbf{x}_{AA}^R &= {}^R\mathcal{A}^{-1}(\mathbf x_{RTB}) = \textrm h^{-1}\circ\textrm f_{AA}^{-1}\circ {}^R\mathcal{T}^{-1}\circ \textrm g^{-1}(\mathbf x_{RTB})\\
  \mathbf{x}_{AA}^R &= {}^R\mathcal{N}^{-1}(\mathbf x_{RTB}) = \textrm h^{-1}\circ\textrm f_{AA}^{-1}\circ {}^R\mathcal{T}_{num}^{-1}\circ \textrm g^{-1}(\mathbf x_{RTB})
\end{aligned}
\end{equation}

The coefficients of the Earth-Moon $L_1$ resonant ER3BP action-angle Hamiltonian are provided in Table \ref{tab:L1EMresonant}. Notice that the resonant coefficients of higher-degree terms are of lower magnitude than those of the Birkhoff ER3BP Hamiltonian. This amounts to a larger radius of validity of the resonant normal form compared to the Birkhoff normal form.

\begin{figure}[htb!]
    \centering
    \subfigure[Earth-Moon\label{subfig:RadiusofValidityEM}]{
        {\includegraphics[width=0.31\textwidth]{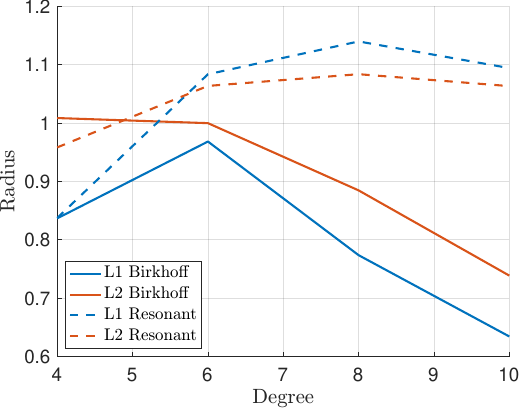}}
        }
    \subfigure[Sun-Earth\label{subfig:RadiusofValiditySE}]{
        {\includegraphics[width=0.31\textwidth]{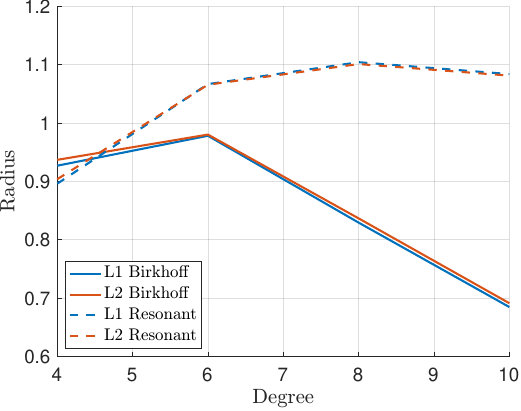}}
        }
    \subfigure[Sun-EM Barycenter\label{subfig:RadiusofValiditySEMB}]{
        {\includegraphics[width=0.31\textwidth]{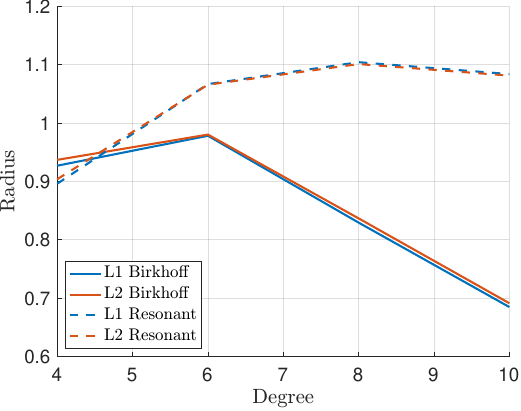}}
        }
        \caption{Radii of validity of the action-angle Hamiltonians for several systems.\label{fig:RadiusofValidity}}
\end{figure}

The radius of validity can be computed from the action-angle Hamiltonian coefficients using the process described in \cite{peterson2024dynamics}. Figure \ref{fig:RadiusofValidity} shows that the radius of validity of both normal forms is greater than those obtained in \cite{paez2021transits,paez2022semi}. As expected, the radius of validity of the resonant normal form is greater than that of the Birkhoff normal form for higher truncation degrees.
\begin{equation}
  r_n = \frac 1 {\sqrt[2n]{\|H_n\|_1}},\quad \|H_n\|_1 = \sum_{\|\mathbf k\|_1=n}|h_{\mathbf k}|,\quad n=1,\dots,\left\lfloor\frac{N}{2}\right\rfloor
\end{equation}
\section{Trajectory Characterization}
The normal forms can be used to characterize five types of center manifolds, along with their stable and unstable manifolds. These include: horizontal Lyapunov tori, vertical Lyapunov tori (often referred to as ``vertical tori''), halo tori, Lissajous trajectories, and quasihalo trajectories. The Birkhoff normal form, with its three constant actions, can describe Lissajous trajectories but not any trajectories within the quasihalo family. The resonant normal form's fluctuating $\hat I_2$ allows for a fifth distinct dynamical behavior which describes quasihalo trajectories. Table \ref{tab:Characterization} summarizes the parameterizations of these center manifold trajectories provided by both the Birkhoff and Resonant normal forms of the ER3BP. Any variable for which an explicit dependence on true anomaly, $f$, is not emphasized is a constant along a member of the respective family of center manifold trajectories.
\begin{table}[ht!]
    \caption{Parameterization of center manifold trajectories as well as their associated invariant manifolds.}\label{tab:Characterization}
    \centering
    \begin{tblr}{Q[c,m,0.15\linewidth]|Q[c,m,0.15\linewidth]|Q[l,m,0.3\linewidth]|Q[l,m,0.3\linewidth]} 
    & &  \SetCell[c=1]{c}{Birkhoff} & \SetCell[c=1]{c}{Resonant} \\ \hline \hline 
    \SetCell[r=5]{c}{\centering Center Manifold \\
    $(\xtil=\pxtil=0)$}&{Lyapunov Trajectory} & \SetCell[c=1]{c}{$\left[0,I_2,0,0,\phi_2(f),0\right]^T$} & \SetCell[c=1]{c}{$\left[0,\hat{I}_2,\hat{I}_3,0,\theta_2(f),\theta_3(f)\right]^T$ \\ $\hat{I}_2 = \hat{I}_3$} \\ \hline
                       &{Vertical Trajectory} & \SetCell[c=1]{c}{$[0,0,I_3,0,0,\phi_3(f)]^T$} & \SetCell[c=1]{c}{$\left[0,0,\hat{I}_3,0,0,\theta_3(f)\right]^T$} \\ \hline 
                       &{Halo Trajectory} & \SetCell[c=1]{c}{$[0,I_2,I_3,0,\phi_2(f),\phi_2(f)\pm \frac{\pi}{2}]^T$ $\phi'_2(f) = \phi'_3(f)$} & \SetCell[c=1]{c}{$\left[0,\hat{I}_2,\hat{I}_3,0,\pm \frac{\pi}{2},\theta_3(f)\right]^T$ \\ $\theta'_2(f)=0$} \\ \hline
                       &{Lissajous Trajectory} & \SetCell[c=1]{c}{$[0,I_2,I_3,0,\phi_2(f),\phi_3(f)]^T$ \\ $\phi_2' \neq \phi_3'$} & \SetCell[c=1]{c}{$\left[0,\hat{I}_2(f),\hat{I}_3,0,\theta_2(f),\theta_3(f)\right]^T$ $\theta_2'(f) > 0\, \forall f$} \\ \hline
&{Quasihalo Trajectory} & \SetCell[c=1]{c}{N/A} & \SetCell[c=1]{c}{$\left[0,\hat{I}_2(f),\hat{I}_3,0,\theta_2(f),\theta_3(f)\right]^T$ $|\theta_2(f) - \theta^*| < \frac{\pi}{2}\, \forall f$, $\theta^* = \pm \frac{\pi}{2}$} \\ \hline \hline
    \SetCell[r=2]{c}{\centering Non-zero Saddle Subspace}
&{Unstable Manifold} & \SetCell[c=1]{c}{$(\xtil,\pxtil)=(\pm \sigma,0)$} &\SetCell[c=1]{c}{$(\xtil,\pxtil)=(\pm \sigma,0)$}  \\ \hline
&{Stable Manifold} &\SetCell[c=1]{c}{$(\xtil,\pxtil)=(0,\pm \sigma)$}  & \SetCell[c=1]{c}{$(\xtil,\pxtil)=(0,\pm \sigma)$} \\  \hline   
    \end{tblr}
\end{table}

Note that although the characterization provided by the ER3BP strongly resembles that of the CR3BP \cite{hunsbergerschwab2025hawaii}, the trajectories of the ER3BP have an additional frequency. Hence, the periodic orbits of the CR3BP become quasiperiodic 2-tori (except in the case of resonant orbits), while the already-quasiperiodic trajectories of the CR3BP become 3-tori.
\subsection{Center Manifold trajectories}
\subsubsection{Lyapunov Trajectories}
\begin{figure}[htb!]
    \centering
\begin{minipage}{0.4\textwidth}
    \centering
    \includegraphics[width=\textwidth]{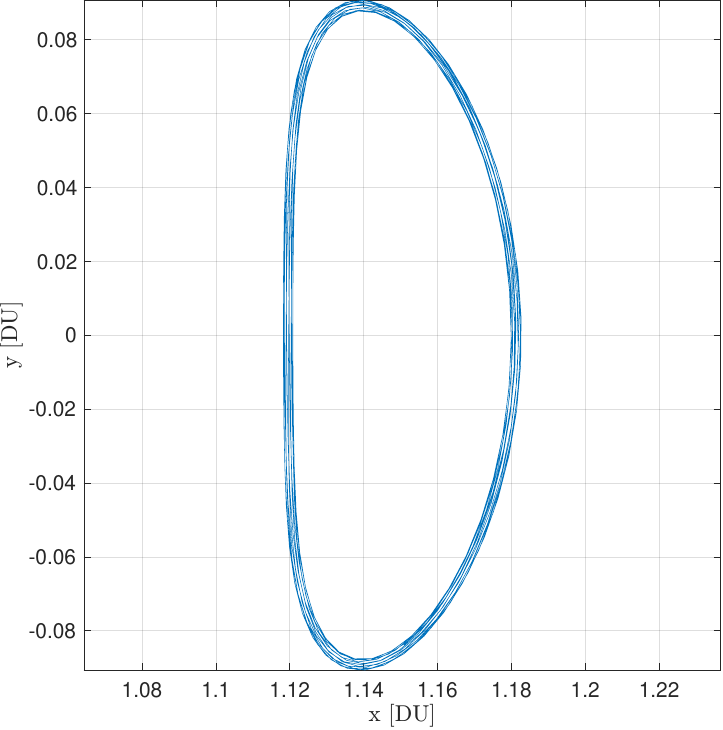} 
\end{minipage}%
\begin{minipage}{0.5\textwidth}
    \centering
    \includegraphics[width=\linewidth]{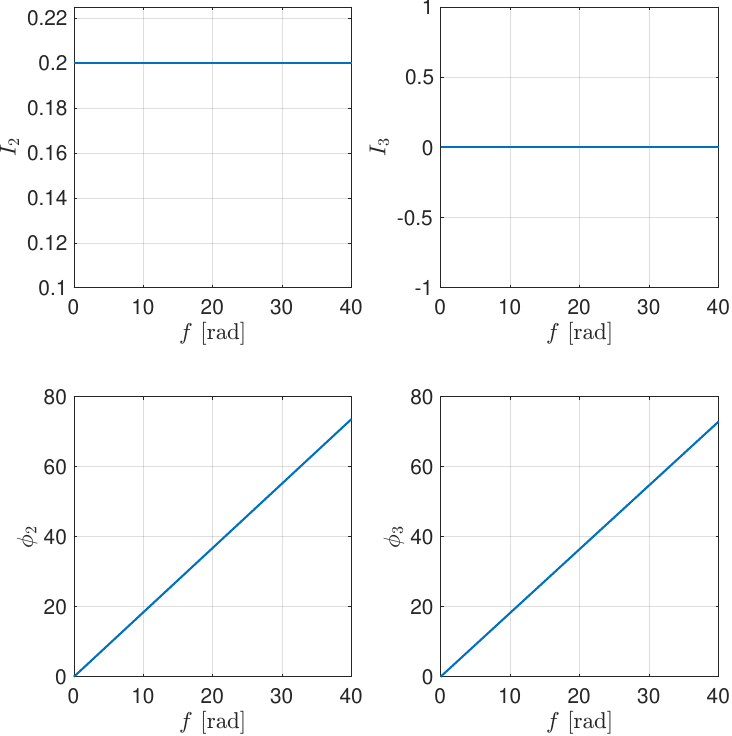}
  \end{minipage}
\caption{Earth-Moon $L_2$ ER3BP Birkhoff normal form horizontal Lyapunov trajectory $(I_2,I_3,\phi_2(0)) = (0.2,0,0)$.\label{fig:L2Lyapunov}}
\end{figure}

The family of Lyapunov tori resides in the $x$-$y$ plane, and can be characterized by both the Birkhoff and resonant normal forms. In fact, it is the only family for which the characterization accuracy is identical between the two models. Since they perform identically, it is logical to use the much simpler Birkhoff normal form. Figure \ref{fig:L2Lyapunov} illustrates the quasiperiodic nature of the Lyapunov trajectories in the ER3BP. Note that although the $\phi_3$ angle is nonzero, the corresponding action $I_3$ is zero, meaning that any value of $\phi_3$ will map to the same point in the rotating-pulsating frame. 

Note that Fig. \ref{fig:L2Lyapunov} displays the Lyapunov trajectory in the rotating-pulsating frame. To get a better idea of the effect that the eccentricity of the system will have on the trajectory, the trajectory can instead be transformed into the non-pulsating frame. Figures \ref{subfig:EMLyapunovrotating} and \ref{subfig:SELyapunovrotating} depict Earth-Moon and Sun-Earth Lyapunov trajectories in their respective non-pulsating frames. The fact that the Sun-Earth Lyapunov trajectory is more elongated may seem counterintuitive given that the Earth-Moon system has a greater eccentricity, but the significantly greater distance between the Sun and the Earth overcomes this effect.

\subsubsection{Vertical Trajectories}
In both normal forms, the family of vertical trajectories is parameterized with a single nonzero action and one linearly varying angle. Although the parameterization is similar, the characterization accuracy of the resonant normal form is better for the family of vertical trajectories. This is because the additional terms kept in the Hamiltonian increase the domain of validity in the out-of-plane direction, as demonstrated in \cite{hunsbergerschwab2025hawaii}. Once again, the nonzero $\theta_2$ can be ignored because $\hat I_2=0$. Figures \ref{subfig:EMverticalrotating} and \ref{subfig:SEverticalrotating} show how the Sun-Earth vertical trajectory is more elongated in the non-pulsating frame than the Earth-Moon trajectory with the same action.
\begin{figure}[htb!]
    \centering
\begin{minipage}{0.3\textwidth}
    \centering
    \includegraphics[width=\textwidth]{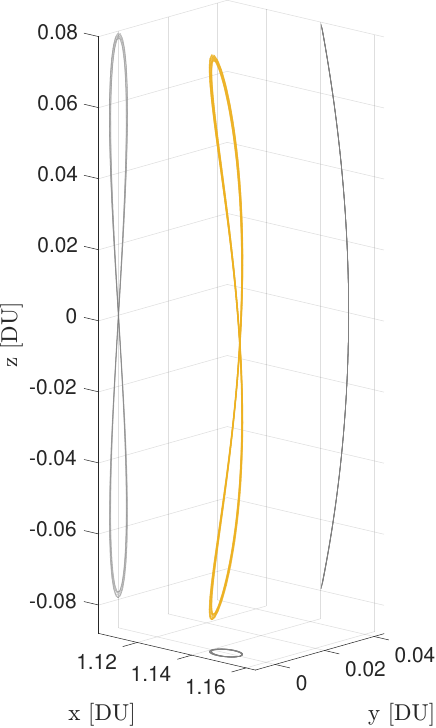} 
\end{minipage}%
\begin{minipage}{0.5\textwidth}
    \centering
    \includegraphics[width=\linewidth]{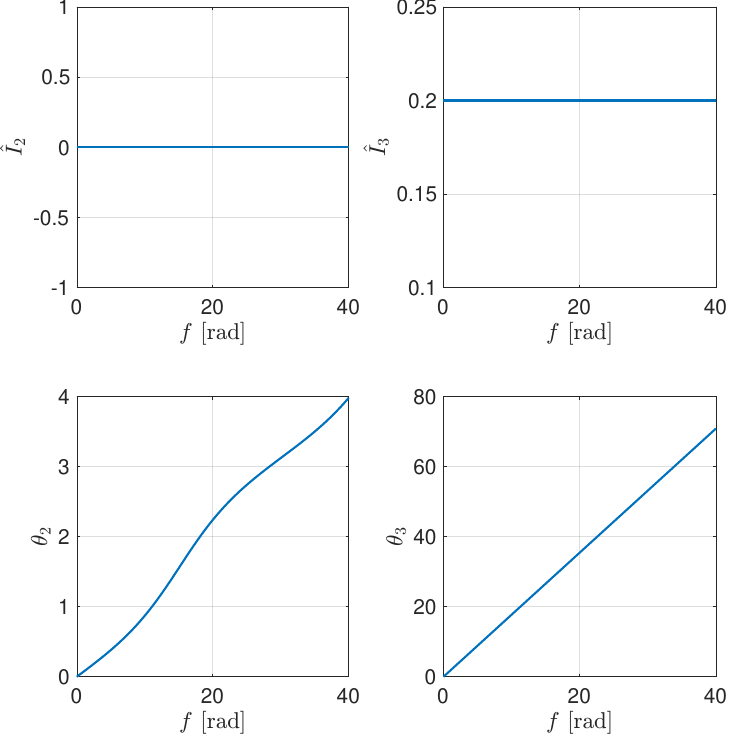}
  \end{minipage}
\caption{Earth-Moon $L_2$ ER3BP resonant normal form vertical Lyapunov trajectory $(\hat I_2,\hat I_3,\theta_3(0)) = (0,0.2,0)$.\label{fig:L2vertical}}
\end{figure}

\subsubsection{Halo Trajectories}
\begin{figure}[htb!]
    \centering
\begin{minipage}{0.5\textwidth}
    \centering
    \includegraphics[width=\textwidth]{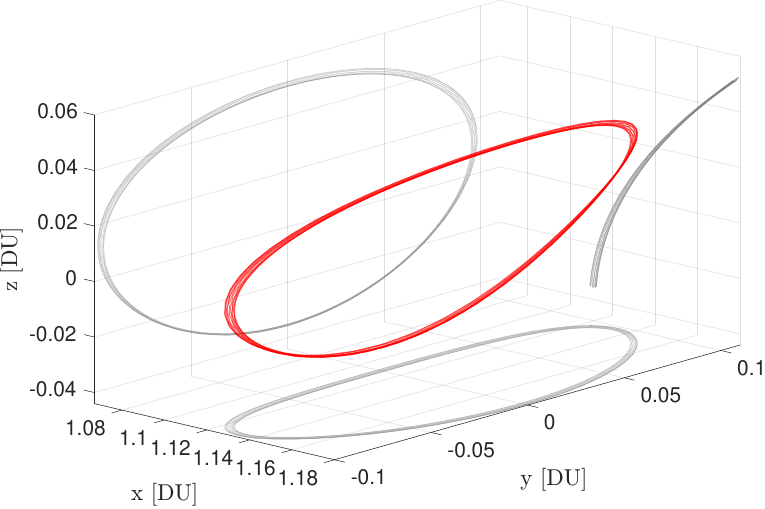} 
\end{minipage}%
\begin{minipage}{0.5\textwidth}
    \centering
    \includegraphics[width=\linewidth]{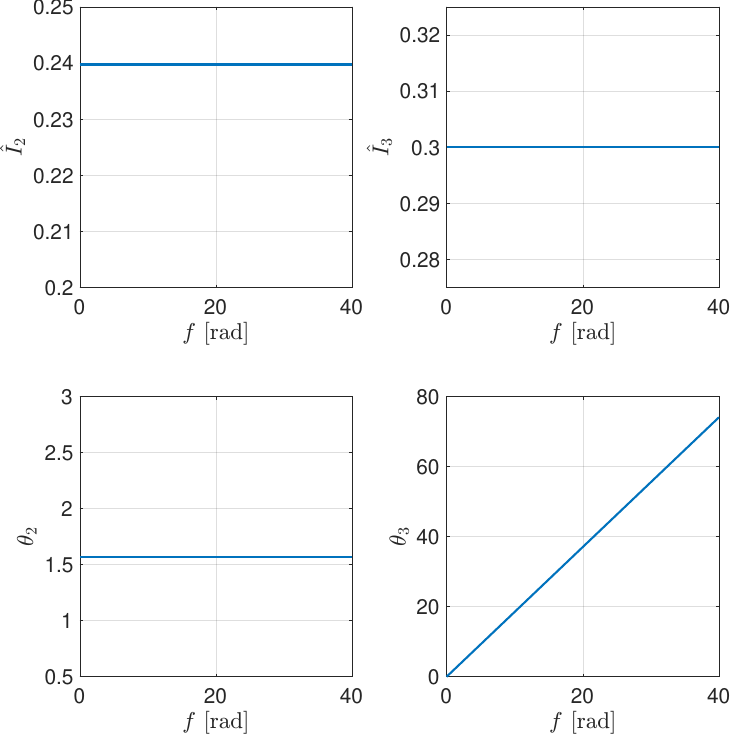}
  \end{minipage}
\caption{Northern Earth-Moon $L_2$ ER3BP resonant normal form halo trajectory $(\hat I_2,\hat I_3,\theta_2(0)) = (0.2398,0.3,\pi/2)$.\label{fig:L2halo}}
\end{figure}

The Birkhoff normal form's inability to accurately characterize trajectories with large out-of-plane motion extends to halo trajectories. Similar to the Birkhoff normal form of the CR3BP, only extremely small halo trajectories can be represented \cite{hunsbergerschwab2025hawaii}. Thus, the resonant ER3BP normal form should be used when working with halo trajectories. An example of a northern Earth-Moon $L_2$ ER3BP resonant normal form halo trajectory is plotted in the rotating-pulsating frame in \ref{fig:L2halo}, along with its corresponding action-angle representation.

Notice that the angle $\theta_2=\frac{\pi}{2}$, implying that the southern family of $L_2$ halo trajectories will have $\theta_2=-\frac{\pi}{2}$. At $L_1$, the signs are flipped, with $\theta_2=\frac{\pi}{2}$ corresponding to the southern family. The sign convention is a result of the non-unique action-angle transformation \ref{eq:fAA}. A different $f_{AA}$ could result in the signs being flipped.

Halo trajectories are fixed points in the $(\hat I_2,\theta_2)$ space, and can be located by finding the actions $\hat I_2$ and $\hat I_3$ that result in the condition $\theta_2'=0$.
\subsubsection{Lissajous Trajectories}

\begin{figure}[htb!]
    \centering
\begin{minipage}{0.5\textwidth}
    \centering
    \includegraphics[width=\textwidth]{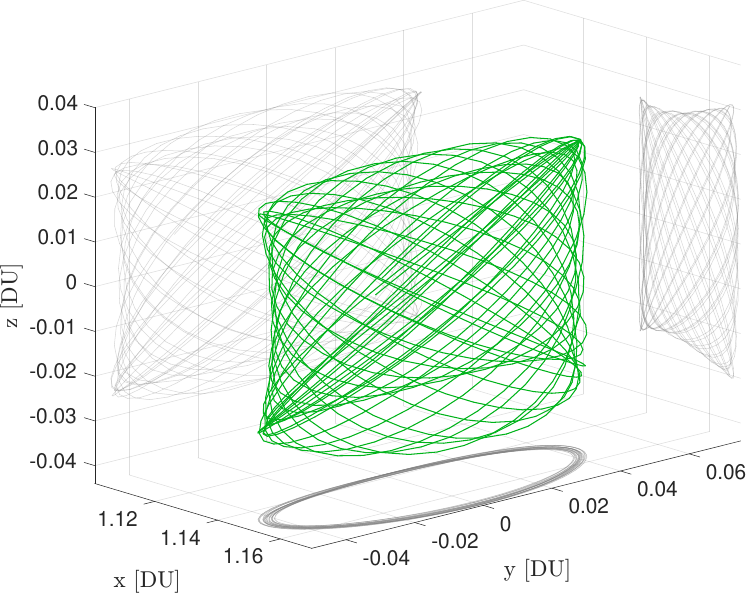} 
\end{minipage}%
\begin{minipage}{0.5\textwidth}
    \centering
    \includegraphics[width=\linewidth]{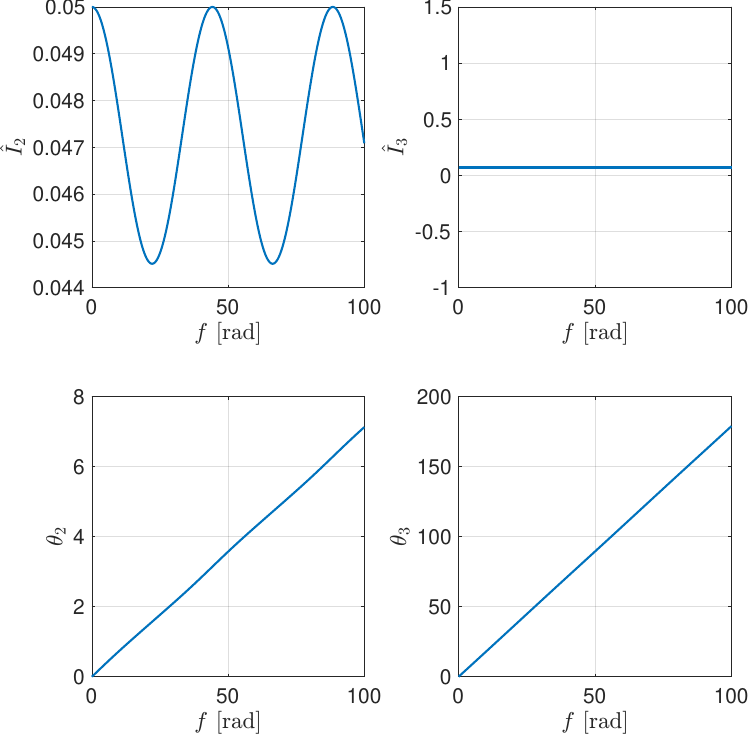}
  \end{minipage}
\caption{Earth-Moon $L_2$ ER3BP resonant normal form Lissajous trajectory $(\hat I_2(0),\hat I_3,\theta_2(0)) = (0.5,0.7,0)$.\label{fig:L2Lissajous}}
\end{figure}

Lissajous trajectories can be represented with both the Birkhoff and Resonant ER3BP normal forms. The Birkhoff normal form describes Lissajous trajectories with constant $I_2$ and $I_3$ actions, making them easier to work with for optimal control purposes. However, just like with all of the other out-of-plane trajectories, the Resonant normal form provides a more accurate approximation of Lissajous trajectories by allowing $\hat I_2$ to oscillate over time. The behavior that separates the Lissajous family from the quasihalo family, as shown in Table \ref{tab:Characterization}, is the monotonic increase in $\theta_2$, illustrated in Fig. \ref{fig:L2Lissajous}. Note that the rate of increase of $\theta_2$ is significantly slower than that of $\theta_3$.

\subsubsection{Quasihalo Trajectories}

\begin{figure}[htb!]
    \centering
\begin{minipage}{0.5\textwidth}
    \centering
    \includegraphics[width=\textwidth]{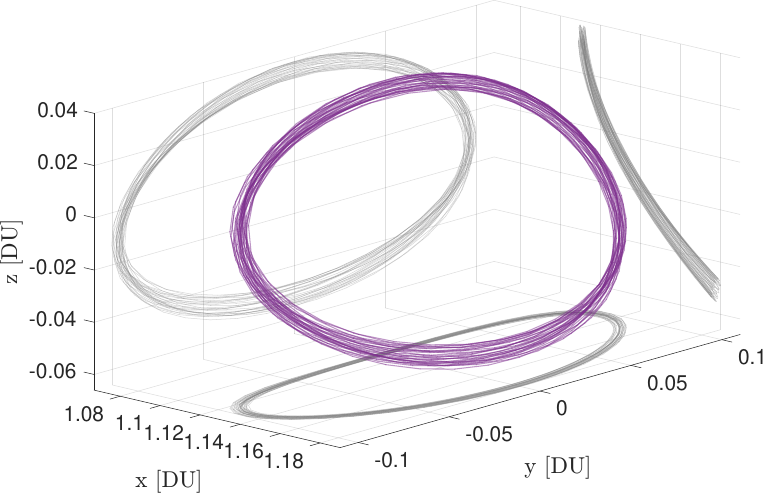} 
\end{minipage}%
\begin{minipage}{0.5\textwidth}
    \centering
    \includegraphics[width=\linewidth]{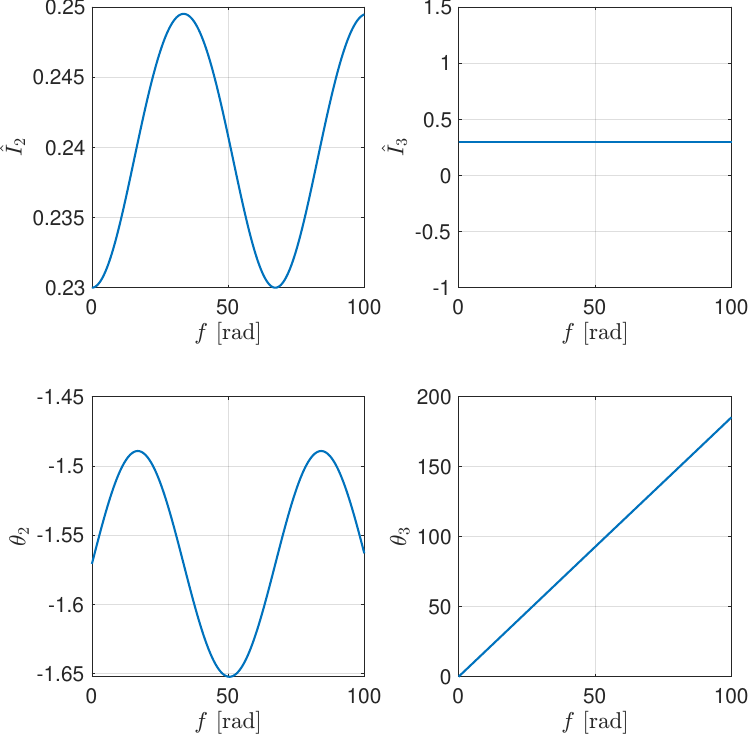}
  \end{minipage}
\caption{Earth-Moon $L_2$ ER3BP resonant normal form quasihalo trajectory $(\hat I_2(0),\hat I_3,\theta_2(0)) = (0.23,0.3,-\pi/2)$.\label{fig:L2quasihalo}}
\end{figure}

The families of quasihalo trajectories cannot be characterized by the Birkhoff normal form. Due to the constant Birkhoff actions and linearly varying angles, for any nonzero actions $I_2$ and $I_3$,the one angle will always increase at a faster rate than the other (except in the case where the two frequencies are identical, corresponding to a halo trajectory). The fluctuation of $\hat I_2$ permitted by the resonant normal form allows for a distinct scenario, one in which the frequencies oscillate about one another, resulting in pure oscillation in $\theta_2$. This behavior corresponds to the quasihalo families of trajectories, as shown in Fig. \ref{fig:L2quasihalo}.

While the winding of the ER3BP Lissajous trajectory appears similar to that of the CR3BP Lissajous trajectories shown in \cite{hunsbergerschwab2025hawaii}, the winding of the ER3BP quasihalo trajectory is visibly different from the even winding of a CR3BP quasihalo trajectory due to the addition of a third frequency.

\subsubsection{Center Manifold Characterization Error}
Although the normal forms are, in theory, capable of characterizing the five aforementioned families of trajectories, there is a decrease in characterization accuracy as the distance from the libration point increases. This error growth can be visualized by selecting a point on a center manifold trajectory within the action-angle space, transforming it into the rotating-pulsating frame with the analytical transformation, $\mathcal A$, and then numerically propagating the state using the full ER3BP equations of motion for one period. As the actions increase, it can be expected that the numerically propagated trajectory will deviate farther from the trajectory obtained through propagation of the initial action-angle state directly within the action-angle space.

\begin{figure}[htb!]
    \centering
    \subfigure[\label{subfig:LyapunovErrorRTB}]{
        {\includegraphics[width=0.35\textwidth]{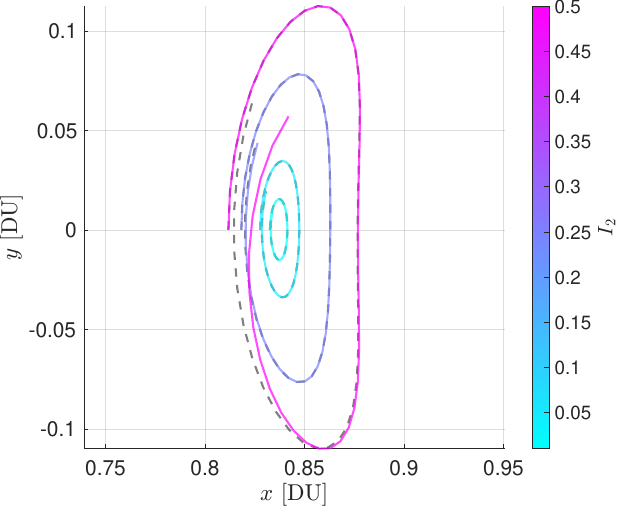}}
        }
    \subfigure[\label{subfig:LyapunovErrorLog}]{
        {\includegraphics[width=0.35\textwidth]{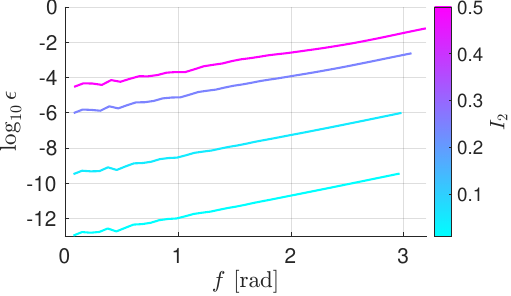}}
        }\\
    \subfigure[\label{subfig:HaloErrorRTB}]{
        {\includegraphics[width=0.35\textwidth]{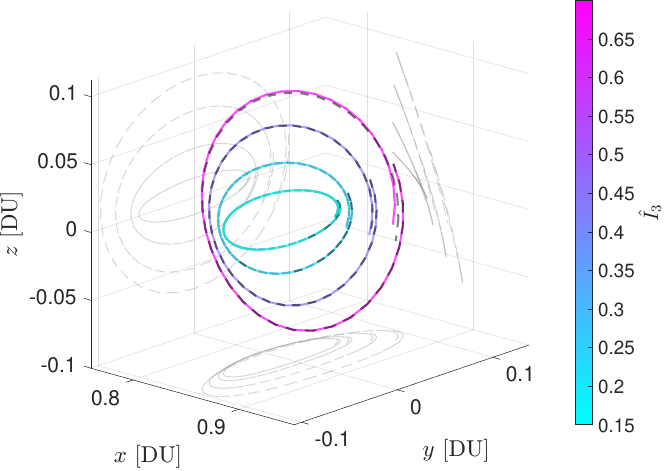}}
        }
    \subfigure[\label{subfig:HaloErrorLog}]{
        {\includegraphics[width=0.35\textwidth]{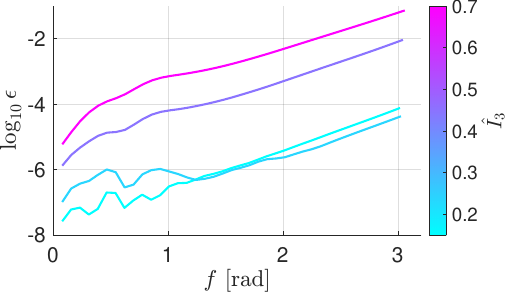}}
        }
        \caption{Difference between propagated states for Earth-Moon $L_1$ Lyapunov and Halo trajectories of varying sizes.\label{fig:PropError}}
\end{figure}

Figure \ref{fig:PropError} confirms that this is the case. Here, the error between the states propagated in the rotating-pulsating and action-angle spaces, $\epsilon$, is simply the distance between the two trajectories number of sampled values of true anomaly.
\begin{equation}\label{eq:epsilon}
  \epsilon_j = \|\mathbf{x}_{RTB}(f_j) - \mathcal A\left(\mathbf x_{AA}(f_j)\right)\|_2
\end{equation}
There is one peculiarity; Fig. \ref{subfig:HaloErrorLog} shows a halo trajectory corresponding to a greater $\hat I_3$ action that experiences a lesser rate of departure than a trajectory with a lower $\hat I_3$ after $f\approx 1.3$. Aside from this case, the general trend shows an increased rate of departure for larger actions, which is the result of a larger initial error. This is to say that the analytical transformation, $\mathcal A$, decreases in accuracy as the actions increase.
In all cases, the initial true anomaly was chosen to be zero. While this choice of initial true anomaly does affect the way in which the trajectories depart, the trends shown are independent of the initial value of $f$. The Birkhoff normal form is used for the Lyapunov case, since the normal forms are equivalent for planar motion, while the resonant normal form is used for the halo case. For the Lyapunov trajectories shown, $\phi_2(0) = 0$, and for the halo trajectories, $\theta_3(0) = \pi$ for the sake of visualization.

The implication of this result is that long-term propagation is not a strength of the normal forms (i.e., one should not propagate a state for a long duration within the action-angle space and expect the corresponding trajectory propagated within the rotating-pulsating frame to remain nearby). If the normal forms are to be leveraged as an initial guess for an algorithm used to obtain numerically center manifold trajectories of the ER3BP, it would likely be necessary to implement a multiple-shooting approach, whereby sampling multiple points along a trajectory in the action-angle space and approximating the trajectory within the rotating-pulsating frame in a piecewise manner would greatly reduce the error at the collocation points.

\begin{figure}[htb!]
    \centering
    \subfigure[$L_1$ Birkhoff\label{subfig:L1Birkhoff}]{
        {\includegraphics[width=0.31\textwidth]{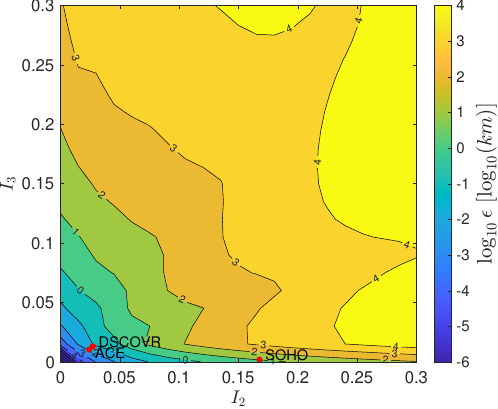}}
        }
    \subfigure[$L_1$ Resonant\label{subfig:L1resonant}]{
        {\includegraphics[width=0.31\textwidth]{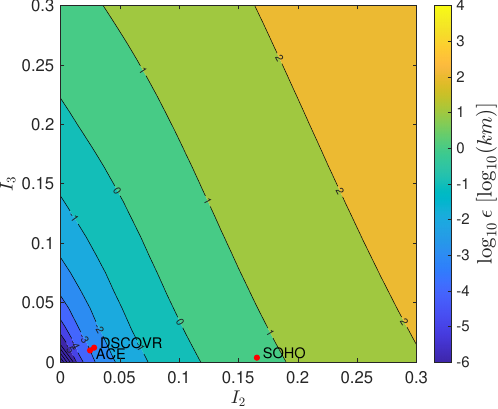}}
        }
    \subfigure[$L_1$ Missions\label{subfig:L1Spacecraft}]{
        {\includegraphics[width=0.31\textwidth]{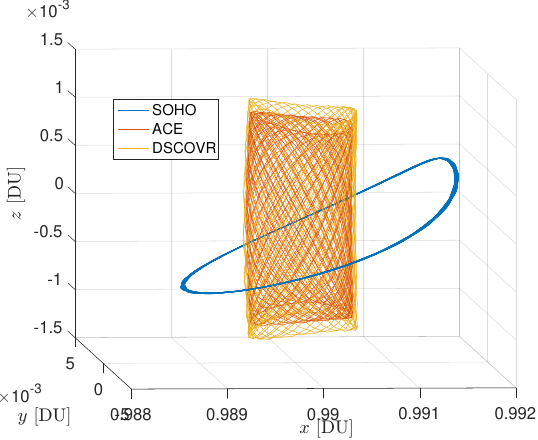}}
        } \\
    \subfigure[$L_2$ Birkhoff\label{subfig:L2Birkhoff}]{
        {\includegraphics[width=0.31\textwidth]{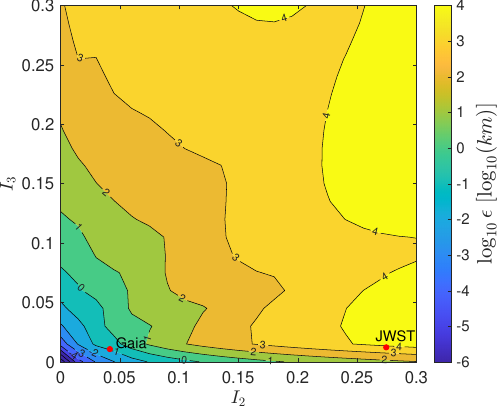}}
        }
    \subfigure[$L_2$ Resonant\label{subfig:L2resonant}]{
        {\includegraphics[width=0.31\textwidth]{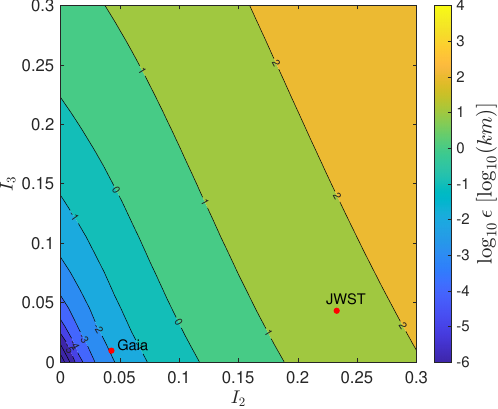}}
        }
    \subfigure[$L_2$ Missions\label{subfig:L2Spacecraft}]{
        {\includegraphics[width=0.25\textwidth]{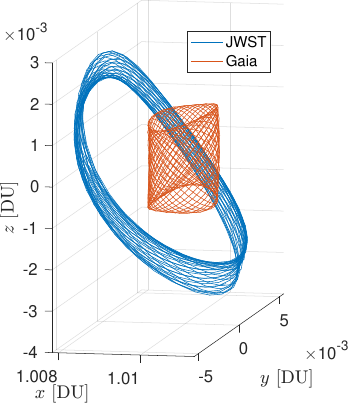}}
        }\hfill

        \caption{Center manifold propagation error for $I_2,I_3 \leq 0.3$, with several Sun-Earth libration point missions plotted for reference.\label{fig:CenterManifoldError}}
\end{figure}

The above analysis is a bit limited-- the Lyapunov and halo families represent only a small subset of the full center manifold. The same error analysis can be performed while sampling across the full $(I_2,I_3)$ space. For this analysis, the error between the two propagated states is evaluated at a fixed true anomaly of $f=1$, in order to reduce the dimensionality of the resulting figures. Results for the Sun-EM barycenter system ($\mu \approx 3.040423\times 10^{-6},e = 0.01671022$) are shown in Fig. \ref{fig:CenterManifoldError} for both $L_1$ and $L_2$, and both the Birkhoff and resonant normal forms. The displayed position errors are in dimensions of kilometers prior to the application of the logarithm.

Figure \ref{fig:CenterManifoldError} makes evident the improved center manifold characterization accuracy of the resonant normal form, which has come at the cost of allowing $\hat I_2$ to fluctuate. The Birkhoff normal form has regions of low error which lie very close to the $I_2$ and $I_3$ axes, but when both actions are nonzero, the propagation error increases quickly. The contour lines in Figs. \ref{subfig:L1resonant} and \ref{subfig:L2resonant} show that the resonant normal form provides low propagation error in between the axes as well. Note that the axes of the resonant normal form figures are the actions $(I_2,I_3)$ rather than $(\hat I_2,\hat I_3)$, in order to better compare the performance of the two normal forms. Figure \ref{fig:CenterManifoldError} also substantiates the earlier claim that the two normal forms provide identical characterization of the planar Lyapunov family, as the intersections of the contour lines with the $I_2$ axis are identical.

To provide an idea of the size of the region contained within the contour plots, the approximate actions of some libration point missions are plotted. These approximations are obtained by taking ephemeris data from \textit{JPL Horizons}, transforming the data into the rotating-pulsating frame as described in Appendix \ref{appendix:Ephemeris}, and then using the inverse numerical transformation, $\mathcal N^{-1}$, to transform the rotating-pulsating states into either the Birkhoff or resonant action-angle spaces. The approximate trajectories of these missions propagated in the resonant action-angle space are shown in Figs. \ref{subfig:L1Spacecraft} and \ref{subfig:L2Spacecraft}. Characterizing these missions with the Birkhoff normal form leads to constant actions, but when using the resonant normal form, trajectories will appear to oscillate on short line segments of constant $\hat I_3 = I_2+I_3$ in the $(I_2,I_3)$ space. The points plotted in Figs. \ref{subfig:L1resonant} and \ref{subfig:L2resonant} are the points on these line segments with maximum $I_2$, since it is observed that these points lie in the region of highest propagation error, and thus provide an upper bound for each trajectory.

\subsection{Saddle Subspace Trajectories}
The normal forms not only provide parameterizations of center manifold trajectories, but also their stable and unstable manifolds. In the saddle subspace spanned by $\tilde x$ and $\tilde p_x$, the center manifold lies at the origin. The $\tilde p_x$-axis is the stable manifold, along which all trajectories approach the origin as $f\to \infty$. The $\tilde x$-axis is the unstable manifold, along which all trajectories move away from the origin at an exponentially increasing rate. All other unstable trajectories will lie in one of the four quadrants-- regions separated by the axes, which correspond to distinct long-term behaviors in the rotating-pulsating state. This subsection will recreate some of the analysis of the saddle subspace covered in \cite{schwab2024characterizing} to demonstrate that the saddle subspace of the ER3BP provides the same insight as the saddle subspace of the CR3BP.

Error analysis of the stable and unstable manifolds can be performed by specifying a center manifold trajectory in the action-angle space, transforming to the corresponding normal form state, and then adding a small step-off in either the stable or unstable direction. From here, the normal form state can be propagated according to Hamilton's equations up until the distance along the invariant manifold reaches a value of one. This is done by propagating backward in time for the stable manifold and forward in time for the unstable manifold. At the same time, the initial normal form state (including the step-off distance) is transformed into the rotating-pulsating frame and propagated according to the ER3BP equations of motion. The propagated normal form state is then transformed into the rotating-pulsating state and compared to the trajectory that was propagated using the ER3BP equations of motion. The error along the manifold is taken to be the distance between the two trajectories.

\begin{figure}[htb!]
    \centering
    \subfigure[\label{subfig:EML2halostable}]{
        {\includegraphics[width=0.4\textwidth]{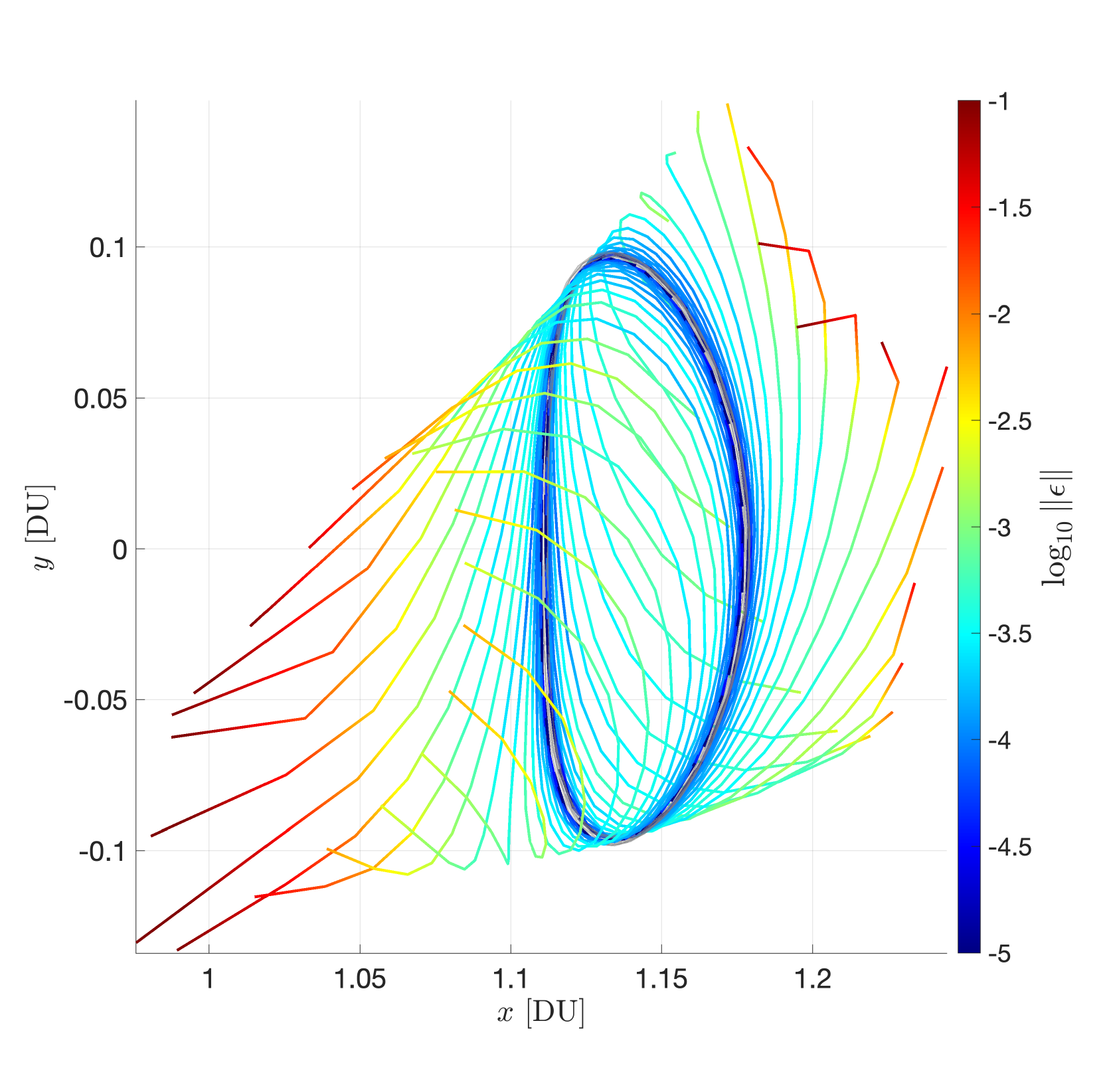}}
        }
    \subfigure[\label{subfig:EML2halounstable}]{
        {\includegraphics[width=0.4\textwidth]{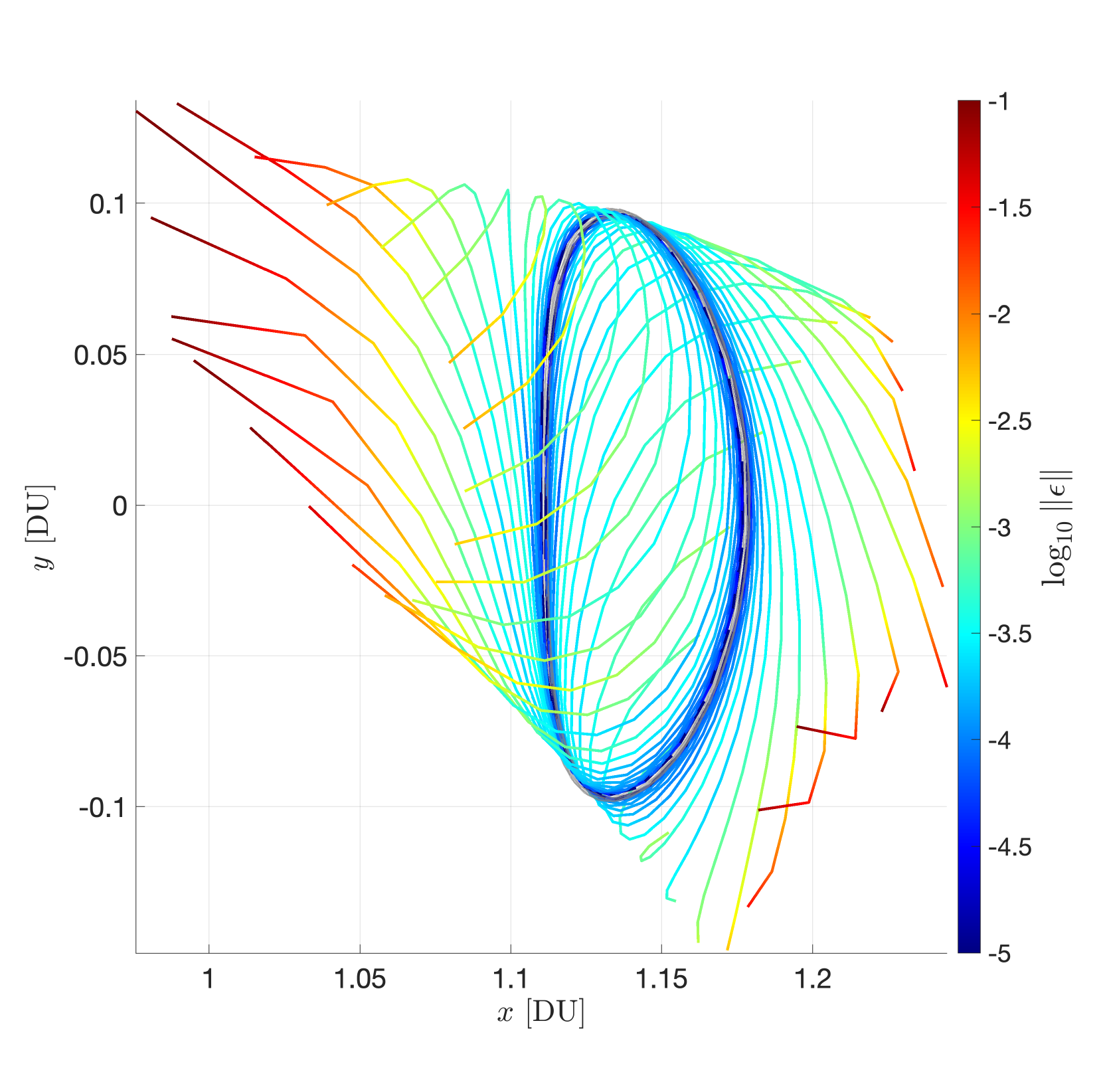}}
        }
        \caption{Invariant manifold error for a northern Earth-Moon L2 resonant normal form halo torus with $(\hat I_2,\hat I_3,\theta_2) = (0.2398,0.3,\pi/2)$.\label{fig:ManifoldError}}
\end{figure}

Since we are working with the ER3BP, the true anomaly must be accounted for. This is done by repeating the aforementioned process for different sampled values of true anomaly, $f_j$. To reduce the size of the resulting figure, the errors for each $f_j$ at the same point along the orbit are averaged and plotted as a single trajectory. Figure \ref{fig:ManifoldError} displays the result of this process for an $L_2$ halo torus with $(\hat I_2,\hat I_3,\theta_2) = (0.2398,0.3,\pi/2)$ (a member of the northern family). The sampled values of true anomaly are $f_j = 0,\frac{\pi}{2},\pi,\frac{3\pi}{2}$, and the step-off distances used are $|\tilde x|,|\tilde p_x| = 0.01$. The error, $\mathbf \epsilon$, is the difference in Cartesian position within the rotating-pulsating frame, in non-dimensional distance units.

Position error along the manifolds increases quickly as they depart from the halo torus. This is to be expected, as the magnitudes of the normal form states before the addition of the step-off distance are already greater than one due to the size of the halo torus. However, the normal form provides accurate characterization of the stable and unstable manifolds within a neighborhood of the halo torus.

\begin{figure}[htb!]
  \centering
  \includegraphics[width=0.5\textwidth]{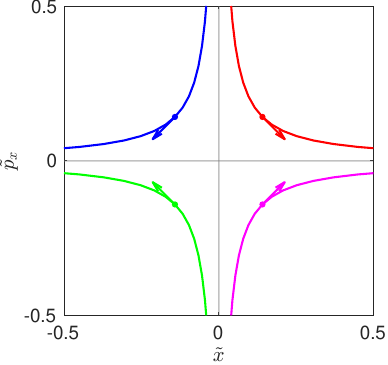}
  \caption{Unstable trajectories in the saddle subspace.\label{fig:Saddle}}
\end{figure}

The normal form is also able to characterize trajectories that do not lie on the $\tilde x$ or $\tilde p_x$ axes. Due to the hyperbolic nature of the saddle subspace, a state propagated in the action-angle or normal form space can never leave the quadrant in which it originates. In this way, the axes separate groups of trajectories that have distinct long-term behaviors. Consider the color-coded unstable trajectories displayed in Fig. \ref{fig:Saddle}, where one has been placed in each quadrant. The initial conditions, represented with points, are obtained by transforming an action-angle state on a Lyapunov trajectory into the normal form space and then adding a step-off of $|\tilde x|,|\tilde p_x| = \sqrt{0.2}$.
\begin{figure}[htb!]
    \centering
    \subfigure[$L_1$ $(I_2=0.2,\phi_2=0)$\label{subfig:L1Saddle0}]{
        {\includegraphics[width=0.35\textwidth]{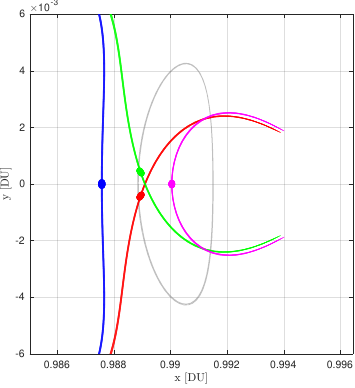}}
        }
    \subfigure[$L_2$ $(I_2=0.2,\phi_2=0)$\label{subfig:L2Saddle}]{
        {\includegraphics[width=0.35\textwidth]{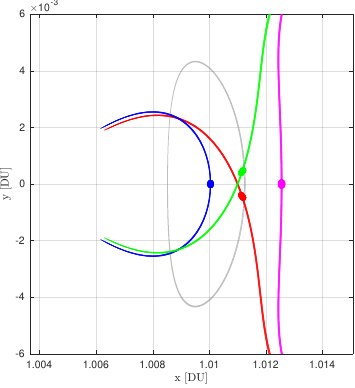}}
        }\\
    \subfigure[$L_1$ $(I_2=0.2,\phi_2=\pi)$\label{subfig:L1Saddle}]{
        {\includegraphics[width=0.35\textwidth]{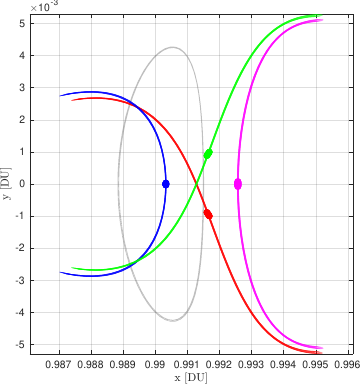}}
        }
    \subfigure[$L_2$ $(I_2=0.2,\phi_2=\pi)$\label{subfig:L2Saddlepi}]{
        {\includegraphics[width=0.35\textwidth]{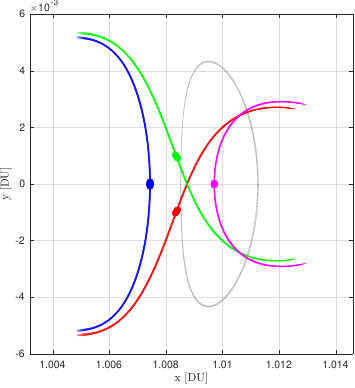}}
        }
        \caption{Saddle subspace quadrants visualized in the Sun-Earth ER3BP rotating-pulsating frame.\label{fig:SaddleRTB}}
\end{figure}

These initial conditions are transformed into the ER3BP rotating-pulsating frame and then propagated forward and backward in true anomaly to visualize where the trajectories originated and where they are headed. Figure \ref{fig:SaddleRTB} shows the saddle trajectories in the rotating-pulsating frame, for both $L_1$ and $L_2$ Lyapunov trajectories and two different points on each trajectory. In all cases, it can be seen that the second quadrant (blue) contains trajectories that approach from the $+x$-direction and are reflected in the $-x$-direction. The fourth quadrant (pink) contains the opposite-- trajectories that approach in the $-x$-direction and are reflected in the $+x$-direction. The other two quadrants contain transit trajectories-- trajectories that approach from one direction and then continue on in that same direction. Trajectories in the first quadrant (red) transit in the $+x$-direction and trajectories in the third quadrant (green) transit in the $-x$-direction.
\begin{figure}[htb!]
  \centering
    \subfigure[Sun-Earth\label{subfig:L1Saddleclose}]{
        {\includegraphics[width=0.335\textwidth]{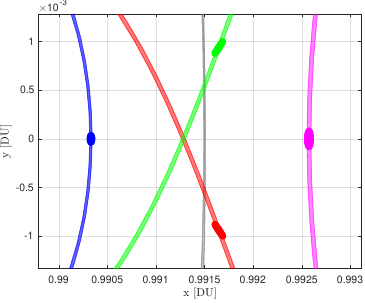}}
        }
    \subfigure[Earth-Moon\label{subfig:L1SaddlecloseEM}]{
        {\includegraphics[width=0.35\textwidth]{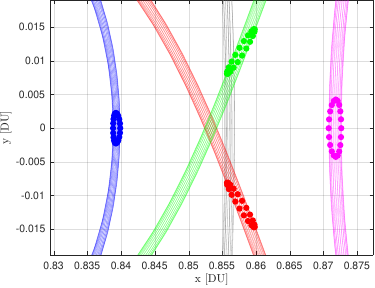}}
        }
  \caption{Close-up view of the saddle trajectories and initial conditions for different systems.\label{fig:Saddleclose}}
\end{figure}

The initial conditions in the saddle subspace will map to different rotating-pulsating states for different values of true anomaly. To illustrate that the long-term behavior of the different quadrants does not depend on true anomaly, the above process is repeated with 20 samples of true anomaly, $f_j$. Figure \ref{fig:Saddleclose} provides a zoomed-in view of the Sun-Earth trajectories in Fig. \ref{fig:SaddleRTB} as well as for the Earth-Moon system, to highlight the fact that a larger eccentricity yields a thicker tube of trajectories.

The saddle subspace provides a powerful tool for characterizing unstable trajectories. Given a rotating-pulsating state, all it takes is a coordinate transformation to determine the long-term behavior of the trajectory without any need for propagation.

\section{Adding Solar Radiation Pressure to the Normal Form}\label{section:SRP}
This section introduces the steps necessary for augmenting the normal forms to account for a simple SRP model, which can be utilized in any system that contains an emitting primary, such as the Sun. While the adjustment is small, the following section will demonstrate that it is necessary for accurately modeling the true dynamics of the Sun-EM barycenter system.

Consider the following simple cannonball model for solar radiation pressure (SRP).
\begin{equation}\label{eq:srppotential}
\mathbf{a}_{SRP} = \frac{\Theta}{r_1^3}\mathbf r_1,\qquad \mathbf r_1 = [(x+\mu),y,z]^T
\end{equation}
where $\Theta$ is a function of several parameters specific to a given satellite:
\begin{equation}
\Theta = \frac{P_0C_RA}{m}.
\end{equation}
Here, $P_0$ is the average solar radiation pressure at a distance of $1\,\text{AU}$ from the Sun, $C_R$ is the coefficient of reflectivity of the satellite, $A$ is the average effective surface area of the satellite, and $m$ is the mass of the satellite. It is important to note that $\Theta$ will be specific to a particular satellite, and thus the normal forms must be derived with a specific satellite in mind.

The addition of SRP results in the following change to the ER3BP equations of motion. It can be thought of as a slight weakening of the Sun's effective gravitational pull, since \eqref{eq:srppotential} has the same form as the potential terms, save for a difference of signs.
\begin{equation}
\begin{aligned}
x''  &= 2y'+\frac{1}{1+e\cos f}\left(x-\frac{(1-\mu-\Theta)(x+\mu)}{r_1^3}-\frac{\mu(x+\mu-1)}{r_2^3}\right)\\
y'' &= -2x'+\frac{1}{1+e\cos f}\left(y-\frac{(1-\mu-\Theta)y}{r_1^3}-\frac{\mu y}{r_2^3}\right)\\
z'' &= -z+\frac{1}{1+e\cos f}\left(z-\frac{(1-\mu-\Theta)z}{r_1^3}-\frac{\mu z}{r_2^3}\right)
\end{aligned}
\end{equation}
Similarly, the ER3BP Hamiltonian can be augmented to account for this new potential term.
\begin{equation}
H = \frac{1}{2}\left(p_x^2+p_y^2+p_z^2\right) - p_yx+p_xy + \frac{1}{1+e\cos f}\left(\frac{1}{2}e\cos f\left(x^2+y^2+z^2 \right)-\frac{1-\mu-\Theta}{r_1}-\frac{\mu}{r_2}\right),
\end{equation}
Only three changes must be made to the ER3BP normal form derivation in order to account for SRP.
\subsection{Libration Point Location}
The locations of the collinear libration points $L_1$ and $L_2$ of the system with SRP can be found by solving the following quintic equation
\begin{equation}
\gamma^5\mp(3-\mu)\gamma^4+(3-2\mu)\gamma^3-(\mu\pm\Theta)\gamma^2\pm2\mu\gamma-\mu=0
\end{equation}
where the upper sign is for $L_1$ and the lower for $L_2$. The added force results in the libration points moving slightly closer to the Sun. This makes intuitive sense as we have effectively weakened the attraction of the Sun, meaning that trajectories can exist slightly closer to the Sun without being pulled down into its gravitational well.
\subsection{Taylor Expansion}
A new approximation for the potential terms is needed.
\begin{equation}
\frac{1-\mu-\Theta}{r_1} + \frac{\mu}{r_2} = \sum_{n=0}^\infty c_n\rho^nP_n\left(\frac{x}{\rho}\right), \quad\rho=\sqrt{x^2+y^2+z^2}
\end{equation}
Once again, the addition of SRP results in a slight adjustment of the $c_n$ coefficients
\begin{equation}
c_n = \frac{1}{\gamma^3}\left[(\pm1)^n\mu + (-1)^n(1-\mu-\Theta)\left(\frac{\gamma}{1\mp\gamma}\right)^{n+1}\right]
\end{equation}
where, as usual, the upper sign is for $L_1$ and the lower for $L_2$.
\subsection{Quadratic Hamiltonian}
To find the symplectic Floquet change of variables, the quadratic part of the Hamiltonian is numerically propagated from $f=0$ to $f=2\pi$.
\begin{equation}
H_2 = \frac{1}{2}\left(p_1^2+p_2^2+p_3^2\right) - p_2q_1+p_1q_2 + \frac{\beta(-2q_1^2+q_2^2+q_3^2)}{1+e\cos f}+\frac{e\cos f(q_1^2+q_2^2+q_3^2)}{2(1+e\cos f)},
\end{equation}
Since the Taylor expansion differs from the non-SRP case, the $\beta$ coefficient must be adjusted to \eqref{eq:betaSRP}.
\begin{equation}\label{eq:betaSRP}
\beta = \frac{1}{2}\left(\frac{1-\mu-\Theta}{|x_L-\mu|^3}+\frac{\mu}{|1-\mu-x_L|^3}\right)
\end{equation}
Note that the characterization properties of the normal forms discussed in the previous section do not change with the addition of SRP.
\section{Comparison to Ephemeris Data}\label{section:ephemeris}
\begin{figure}[htb!]
    \centering
    \subfigure[\label{subfig:JWSTSunCentered}]{
        {\includegraphics[width=0.48\textwidth]{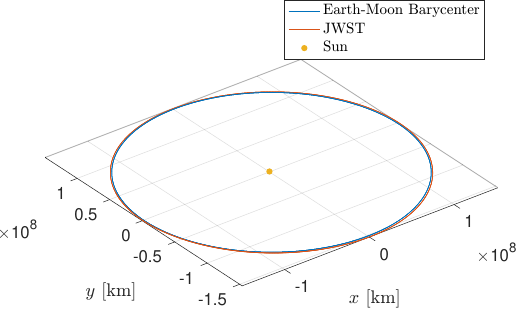}}
        }
    \subfigure[\label{subfig:JWSTER3BPframe}]{
        {\includegraphics[width=0.48\textwidth]{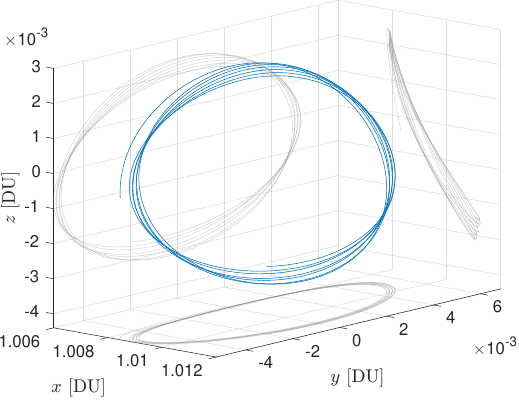}}
        }
        \caption{JWST ephemeris data from 27 Feb 2022 to 7 July 2025 in the Sun-centered frame (left) and rotating-pulsating frame (right).\label{fig:JWST}}
\end{figure}

This section demonstrates the increase in fidelity that comes with transitioning from the CR3BP to the ER3BP and seeks to justify the inclusion of solar radiation pressure within the normal form, all through normal form analysis of ephemeris data obtained from \textit{JPL Horizons}. To begin, consider the James Webb Space Telescope (JWST). The JWST is in a southern $L_2$ Sun-EM barycenter quasihalo trajectory-- one with a maximum $z$-amplitude of about $400{}000\,\textrm{km}$. Figure \ref{subfig:JWSTSunCentered} displays the positions of the JWST and the Earth-Moon barycenter in the Sun-centered frame from 27 Feb 2022 to 1 July 2025, recorded every 24 hours. The Sun-centered ephemeris data can be converted into the rotating-pulsating frame through the steps outlined in  Appendix \ref{appendix:Ephemeris}. Figure \ref{subfig:JWSTER3BPframe} shows the result of this transformation; the JWSt ephemeris data does indeed resemble a southern quasihalo trajectory. The start date of the ephemeris data was selected so that part of the insertion onto the quasihalo torus would be captured, which is visible in Fig. \ref{subfig:JWSTER3BPframe}.
\begin{figure}[htb!]
    \centering
    \subfigure[CR3BP actions\label{subfig:JWSTCR3BPactions}]{
        {\includegraphics[width=0.9\textwidth]{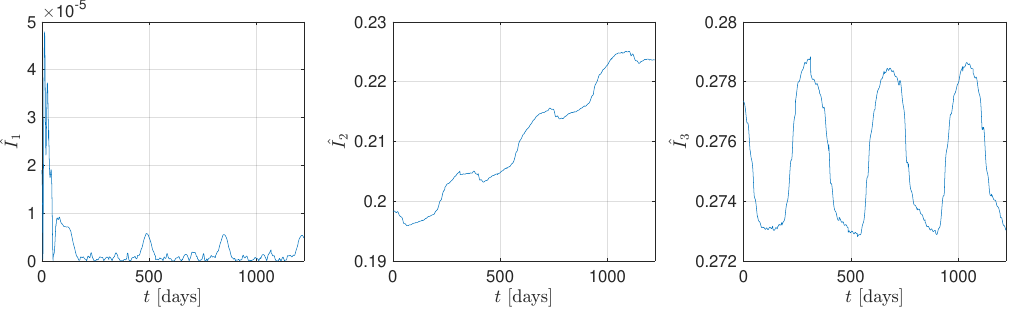}}
        }\\
    \subfigure[CR3BP angles\label{subfig:JWSTCR3BPangles}]{
        {\includegraphics[width=0.6\textwidth]{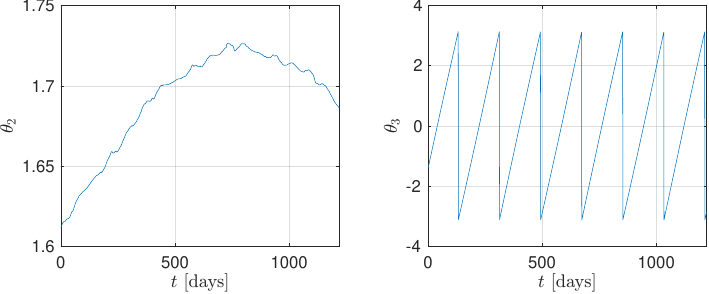}}
        }
        \caption{JWST ephemeris data transformed into the resonant CR3BP action-angle variables.\label{fig:JWSTCR3BP}}
\end{figure}

Let us begin by transforming the rotating-pulsating states into the resonant normal form and action-angle variables of the lower-fidelity CR3BP using the analytical transformation ${}^R\mathcal{A}^{-1}:\mathbf x_{RTB}\to\mathbf x_{AA}^R$. Figure \ref{fig:JWSTCR3BP} shows that all three actions experience noticeable fluctuations with a period of about one year. The unstable action in particular, $\hat I_1$, appears to increase when the Earth-Moon barycenter is near apoapsis in its orbit around the Sun.

\begin{figure}[htb!]
  \centering
    \subfigure[\label{subfig:JWSTCR3BPsaddle}]{
        {\includegraphics[width=0.45\textwidth]{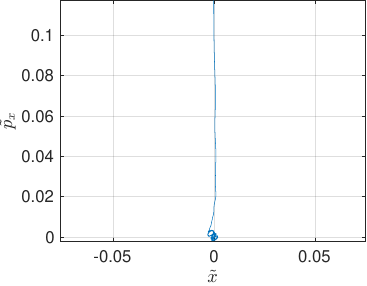}}
        }
    \subfigure[\label{subfig:JWSTCR3BPsaddleclose}]{
        {\includegraphics[width=0.3\textwidth]{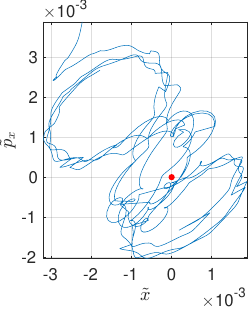}}
        }
  \caption{JWST ephemeris data projected onto the saddle subspace of the resonant CR3BP normal form.\label{JWSTCR3BPsaddle}}
\end{figure}

Further analysis of the saddle subspace, shown in Fig. \ref{JWSTCR3BPsaddle}, implies that the center manifold of the true system is moving around in a periodic manner relative to the center manifold of the CR3BP (indicated in red). This is to be expected, as the CR3BP does not account for the eccentricity of the Earth-Moon barycenter's true orbit. 

\begin{figure}[htb!]
    \centering
    \subfigure[ER3BP actions\label{subfig:JWSTER3BPactions}]{
        {\includegraphics[width=0.9\textwidth]{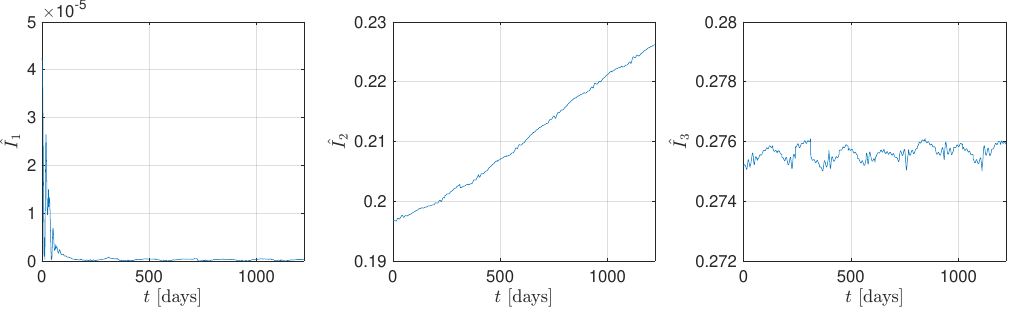}}
        }\\
    \subfigure[ER3BP angles\label{subfig:JWSTER3BPangles}]{
        {\includegraphics[width=0.6\textwidth]{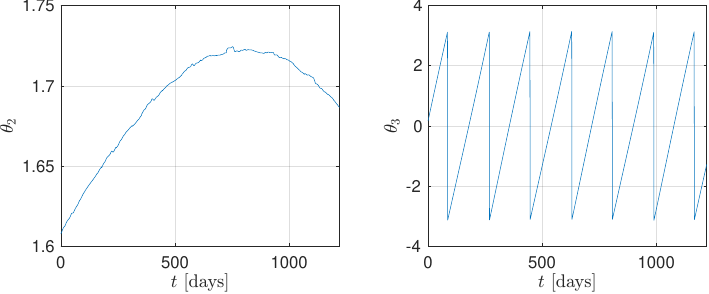}}
        }
        \caption{JWST ephemeris data transformed into the resonant ER3BP action-angle variables.\label{fig:JWSTER3BP}}
\end{figure}

\begin{figure}[htb!]
    \centering
    \subfigure[\label{subfig:JWSTER3BPsaddle}]{
        {\includegraphics[width=0.45\textwidth]{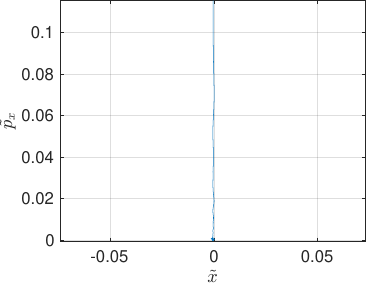}}
        }
    \subfigure[\label{subfig:JWSTER3BPsaddleclose}]{
        {\includegraphics[width=0.3\textwidth]{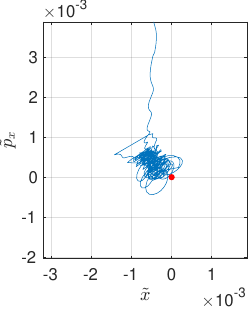}}
        }
        \caption{JWST ephemeris data projected onto the saddle subspace of the resonant ER3BP normal form without SRP.\label{fig:JWSTER3BPsaddle}}
\end{figure}

The same ephemeris data is then transformed into the ER3BP normal form and action-angle coordinates, with the results shown in Fig. \ref{fig:JWSTER3BP} and \ref{fig:JWSTER3BPsaddle}. The ER3BP normal form's abilitiy to characterize the JWST's ephemeris data is noticeably better than that of the CR3BP normal form. The unstable action, as seen in the leftmost plot in Fig. \ref{subfig:JWSTER3BPactions}, is on the order of $10^{-5}$ during the insertion onto the quasihalo trajectory, and then remains on the order of $10^{-7}$ during the JWST's station-keeping phase, with much smaller fluctuations. The other actions behave in a manner consistent with quasihalo trajectories, with $\hat I_2$ increasing and $\hat I_3$ experiencing small fluctuations but remaining fairly constant. The angle $\theta_2$ appears to be oscillating about $\pm\frac{\pi}{2}$, as expected.

Figure \ref{subfig:JWSTER3BPsaddle} shows the ephemeris data projected onto the saddle subspace of the resonant ER3BP normal form. The insertion along the stable manifold is visible, after which the JWST remains close to the origin of the saddle subspace-- where the center manifold of the normal form lies. However, it is important to note that the JWST does not oscillate exactly around the origin of the saddle subspace, but rather almost entirely in the second quadrant, as shown in Fig. \ref{subfig:JWSTER3BPsaddleclose}.

It is helpful to recall that the quadrants of the saddle subspace separate trajectories with distinct long-term behaviors. From Figs. \ref{fig:Saddle} and \ref{fig:SaddleRTB}, the second quadrant contains trajectories that come from and return to the Sun. In other words, the resonant ER3BP normal form believes that at nearly every instant, the JWST should depart in the direction of the Sun. We know that it is not the case; in reality, the JWST remains near $L_2$ in its intended quasihalo trajectory. This means that in the true Sun-EM barycenter system, there is a persistent perturbing force in the positive $x$-direction that the ER3BP normal form does not account for.

\begin{figure}[htb!]
  \centering
    \subfigure[SOHO ($L_1$)\label{subfig:SOHO}]{
    \begin{minipage}{0.31\linewidth}
      \centering
      \vspace{1cm}

      \includegraphics[width=1\linewidth]{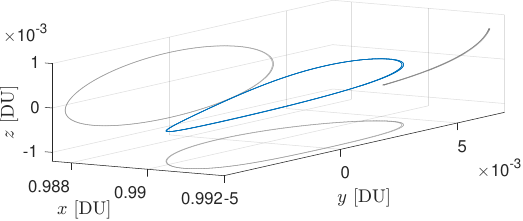}

      \vspace{0.8cm}

      \includegraphics[width=1\linewidth]{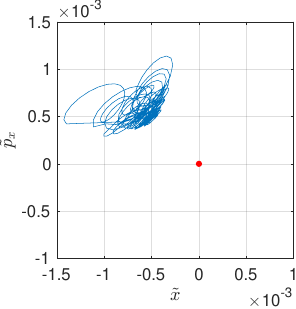}
    \end{minipage}}
    \subfigure[Gaia ($L_2$)\label{subfig:Gaia}]{
  \begin{minipage}{0.31\linewidth}
        \includegraphics[width=1\linewidth]{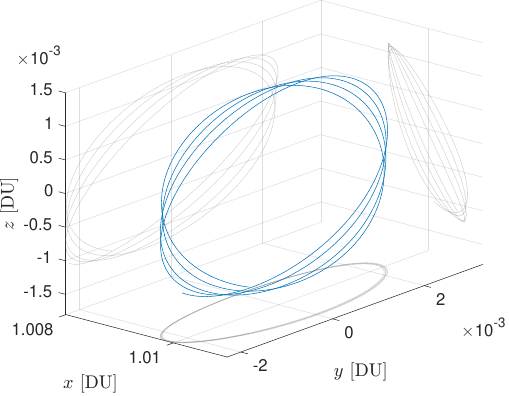}
        \\
        \includegraphics[width=1\linewidth]{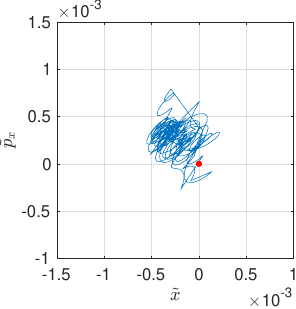}
    \end{minipage}}
    \subfigure[ACE ($L_1$)\label{subfig:ACE}]{
  \begin{minipage}{0.31\linewidth}
        \includegraphics[width=1\linewidth]{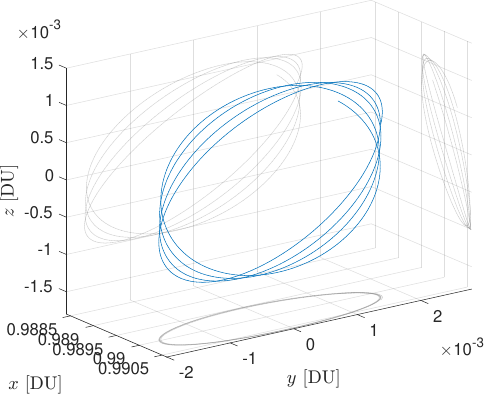}
        \\
        \includegraphics[width=1\linewidth]{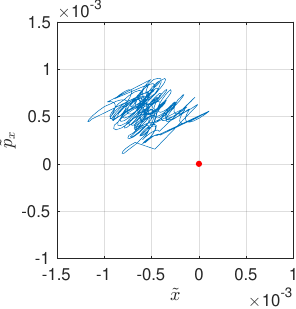}
    \end{minipage}}
\caption{Ephemerides of other missions displayed in the rotating-pulsating frame and projected onto the saddle subspace of the resonant ER3BP normal form without SRP (8 Jan 2016 to 8 Jan 2018).\label{fig:saddlemissions}}
\end{figure}

To verify that this is not an anomaly, the ephemeris data of three other missions is taken and projected into the resonant ER3BP normal form saddle subspace. One other $L_2$ mission is analyzed, Gaia, along with two $L_1$ missions, SOHO and ACE. The trajectories of these three missions are displayed in the rotating-pulsating frame in Fig. \ref{fig:saddlemissions} along with their corresponding projections onto the saddle subspace. The start and end dates of the ephemeris data are held constant for all three missions: 8 Jan 2016 to 8 Jan 2018, with position and velocity recorded every 24 hours. It is clear that the displacement into the second quadrant is not unique to the JWST. Indeed, all three other missions indicate that there is a consistent outward force originating at the Sun. 

\begin{figure}[htb!]
    \centering
    \subfigure[ER3BP + SRP actions\label{subfig:JWSTER3BPactionsSRP}]{
        {\includegraphics[width=0.9\textwidth]{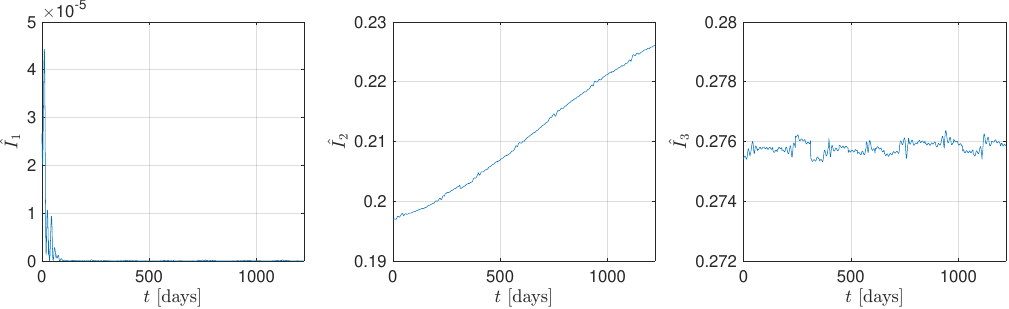}}
        }\\
    \subfigure[ER3BP + SRP angles\label{subfig:JWSTER3BPanglesSRP}]{
        {\includegraphics[width=0.6\textwidth]{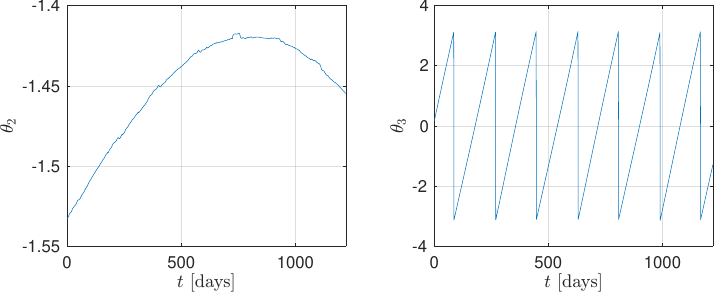}}
        }
        \caption{JWST ephemeris data transformed into the resonant ER3BP including SRP action-angle variables.\label{fig:JWSTER3BPSRP}}
\end{figure}

\begin{figure}[htb!]
    \centering
    \subfigure[\label{subfig:JWSTER3BPsaddleSRP}]{
        {\includegraphics[width=0.45\textwidth]{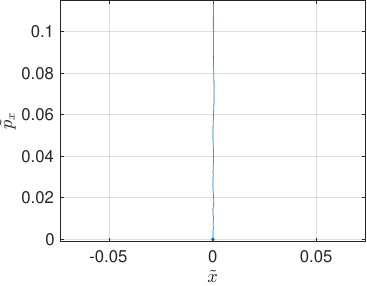}}
        }
    \subfigure[\label{subfig:JWSTER3BPsaddlecloseSRP}]{
        {\includegraphics[width=0.3\textwidth]{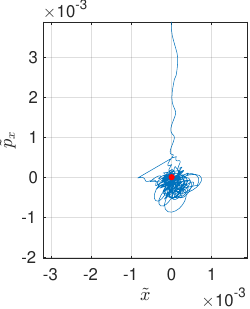}}
        }
        \caption{JWST ephemeris data projected onto the saddle subspace of the resonant ER3BP normal form including SRP.\label{fig:JWSTER3BPsaddleSRP}}
\end{figure}

The most likely candidate for this perturbation is solar radiation pressure. To verify that this is the case, the resonant ER3BP normal form including SRP is generated for the JWST using the steps discussed in Section \ref{section:SRP}.
For the JWST, the following approximate values are used \cite{dichmann2014stationkeeping}.
\begin{equation}
P_0=4.53\times10^{-6}\,\frac{\text N}{\text m^2},\quad C_R = 1.811,\quad A=140\,\text{m}^2,\quad m = 6500\, \text{kg}
\end{equation}
For the JWST and the Sun-EM barycenter system, careful conversion to nondimensional units yields a value of $\Theta \approx 2.97967\times 10^{-5}$. The coefficients of the resonant ER3BP action-angle Hamiltonian about $L_2$ are included in \ref{tab:resonantL2srp}. The JWST ephemeris data is then transformed into the normal form and action-angle variables for the resonant ER3BP normal form including SRP, with the results displayed in Figs. \ref{fig:JWSTER3BPSRP} and \ref{fig:JWSTER3BPsaddleSRP}. Th unstable action is still on the the order of $10^{-5}$ during the insertion due to oscillations about the $\tilde p_x$-axis (shown in Fig. \ref{subfig:JWSTER3BPsaddlecloseSRP}), however when SRP is accounted for, $\hat I_1$ experiences smaller oscillations during the stationkeeping phase, on the order of $10^{-8}$. Similarly, the other actions experience smaller fluctuations than when SRP is not accounted for.

The reason for the smaller magnitude of $\hat I_1$ is obvious upon inspecting Fig. \ref{subfig:JWSTER3BPsaddlecloseSRP}. Including SRP within the ER3BP normal form shifts the origin of the saddle subspace closer to the center of the projected data from the station-keeping phase of the mission.
The significance of this result is that the center manifold resonant ER3BP normal form including SRP now lies in a feasible location. Any optimal control schemes leveraging the normal forms can be justified, because driving the unstable $\tilde x$ component to zero will produce a state quite close to the true stable manifold of the full ephemeris.

It is expected that similar analysis could be repeated with the three other missions shown in Fig. \ref{fig:saddlemissions}, given the spacecraft-specific parameters necessary to compute the nondimensional SRP parameter, $\Theta$.

The results of this analysis should be taken with a grain of salt, as it is not explicitly stated that the target of the JWST's station-keeping is to remain on a specific quasihalo trajectory. Rather, it is stated in \cite{dichmann2014stationkeeping} that maneuvers are performed such that the $x$-component of velocity is zero at a future crossing of the $x$-$z$-plane. Thus, additional conclusions cannot be drawn from the oscillation in $\hat I_3$ shown in Fig. \ref{subfig:JWSTER3BPactionsSRP}. It is not known if $\hat I_3$ is meant to remain near a fixed value, or if it is allowed to gradually change throughout the mission to minimize $\Delta \mathbf v$.

\section{Conclusion}
 It was shown that ER3BP normal forms have reasonable radii of validity and provide trajectory parameterization capabilities that are nearly identical to that of the CR3BP normal forms, the only notable difference being that the true anomaly must be specified when transforming from the normal form and action angle spaces to the rotating-pulsating frame. Additionally, methodology for including a simple cannonball model for solar radiation pressure in the normal forms of the elliptic problem was presented in this paper. Preliminary analysis of the ephemerides of multiple satellites projected into the normal form saddle subspace was used to first demonstrate that the ER3BP provides a more accurate approximation of the dynamics at the libration points of the Sun-EM barycenter system than the CR3BP. Then, additional analysis within the saddle subspace indicated that the normal form model (and by extension the ER3BP itself) failed to capture a constant perturbing force directed outward from the Sun-- solar radiation pressure. Finally, it was shown that augmenting the Hamiltonian to include solar radiation pressure shifted the saddle subspace projections of the ephemeris data of the JWST closer to the origin of the saddle subspace and resulted in overall smaller fluctuations in the actions, implying that solar radiation pressure was indeed the source of the perturbation. It is expected that similar analysis could be performed with the other libration point missions, given their surface area and reflectivity parameters.  The inclusion of solar radiation pressure requires only a few small adjustments to the standard ER3BP normal form derivation, but yields a more accurate dynamical model for the Sun-EM barycenter system.
\section*{Acknowledgments}
This material is based upon work supported jointly by the AFOSR grant FA9550-23-1-0512 and
the Penn State Applied Research Lab (ARL) Walker Assistantship program.
\appendix
\section{Converting from Ephemeris to the Rotating-Pulsating Frame}\label{appendix:Ephemeris}

This section deals with the conversion between a primary-centered frame and the instantaneous rotating-pulsating frame necessary for working with the ER3BP.
\subsection{Position and Velocity}
The process can be divided into two main steps:
 Convert the state into the rotating (but not pulsating) frame, and then
transform the state from the rotating frame into the rotating-pulsating frame.
More specifically, this section will address the processing of Ephemeris data, which has the larger primary as the center of the original coordinate system. Note that a different data format will result in a slightly different process.

The following parameters are used to obtain the results displayed in Section \ref{section:ephemeris} \cite{park2021jpl}.
\begin{equation}\label{eq:Gmvals}
  \begin{aligned}
    Gm_S &= 1.32712440041279419\times10^{11}\,\,\text{km}^3\text s^{-2}\\
    Gm_E &= 398600.435507\,\,\text{km}^3\text s^{-2}\\
    Gm_M &= 4902.800118\,\,\text{km}^3\text s^{-2}
\end{aligned}
\end{equation}
Consider the case in which the ephemeris data has the Sun at the origin, which is to say that
\begin{equation}
\mathbf{r}_S = \mathbf 0,\quad \mathbf v_S = \mathbf{0}.
\end{equation}
The position and velocity of the Earth at an epoch will be given as $\mathbf r_{EM},\mathbf v_{EM} \in\mathbb R^3$, and position and velocity vectors of the satellite of interest at the same epoch will be denoted $\mathbf r,\mathbf v\in \mathbb R^3$.
To begin, we calculate the angular momentum vector of the Earth as well as its magnitude.
\begin{equation}
\mathbf h = \mathbf r_{EM}\times\mathbf v_{EM},\quad h=|\mathbf h|
\end{equation}
The angular velocity vector of the Earth-Moon barycenter with respect to the Sun is then
\begin{equation}
\boldsymbol{\omega} = \frac{\mathbf h}{r_{EM}^2}
\end{equation}
where $r_E = |\mathbf r_{EM}|$. The time derivative of the Earth-Moon system's true anomaly at this particular epoch is simply the magnitude of the angular velocity vector.
\begin{equation}
\dot{f} = |\boldsymbol \omega|
\end{equation}
In the ER3BP (and the CR3BP, for that matter), we will need to shift the origin to the barycenter of the system, which can be done by scaling the vector pointing from the Sun to the Earth-Moon barycenter by a factor of $\mu$
\begin{equation}
\mathbf r_b = \mu\mathbf r_{EM}
\end{equation}
where
\begin{equation}
\mu = \frac{m_E+m_M}{m_S+m_E+m_M}
\end{equation}
is the same $\mu$ that appears in the ER3BP (and CR3BP) equations of motion.
The velocity of the system's barycenter is the following.
\begin{equation}
  \mathbf v_b = \mu \mathbf v_{EM}
\end{equation}
With the position and velocity of the barycenter defined, we can define a new frame with its origin at the barycenter and convert the satellite's position and velocity into this frame.
\begin{equation}
  \begin{aligned}
    {}^B\mathbf r &= \mathbf{r}-\mathbf r_b\\
    {}^B\mathbf v &= \mathbf v-\mathbf v_b
\end{aligned}
\end{equation}
The superscript $B$ stands for ``barycenter'' and will be used to avoid confusion as we will need to step through a few different frames.
Next, we transform the position and velocity into the rotating frame, denoted by a superscript $R$.
To do so, we must first define the three unit vectors of the rotating frame. The $\hat{\mathbf x}$ vector points from the Sun to the Earth-Moon barycenter, the $\hat{\mathbf z}$ vector points in the direction of the angular momentum, and $\hat{\mathbf y}$ is defined such that it completes the right-handed coordinate frame.
\begin{equation}
  \hat{\mathbf{x}} = \frac{\mathbf r_{EM}}{r_{EM}},\quad  \hat{\mathbf z} = \frac{\mathbf h}{h},\quad \hat{\mathbf y} = \hat{\mathbf z}\times \hat{\mathbf x}
\end{equation}
These unit vectors are used to define a direction cosine matrix (DCM) relating the $R$ and $B$ frames, which can be used directly to find the satellite's position in the rotating frame.
\begin{equation}
{}^R \mathbf r = \begin{bmatrix}\hat{\mathbf x}^T\\\hat{\mathbf y}^T\\\hat{\mathbf z}^T\end{bmatrix}{}^B\mathbf r
\end{equation}
The satellite's velocity in the rotating frame can be calculated using the DCM in addition to the transport theorem.
\begin{equation}
{}^R \mathbf v = \begin{bmatrix}\hat{\mathbf x}^T\\\hat{\mathbf y}^T\\\hat{\mathbf z}^T\end{bmatrix}\left({}^B\mathbf v - \boldsymbol\omega \times{}^B\mathbf r\right)
\end{equation}
Let the coordinates and velocity components of the satellite in the rotating frame be denoted as follows
\begin{equation*}
{}^R\mathbf r = [X,Y,Z]^T,\quad {}^R\mathbf v=[\dot X,\dot Y,\dot Z]^T,
\end{equation*}
and let the pulsating-rotating coordinates and their derivatives with respect to $f$ be
\begin{equation*}
{}^P\mathbf r = [x,y,z]^T,\quad {}^P\mathbf v=[ x', y', z']^T.
\end{equation*}
We have the following relations between the rotating coordinates and the rotating-pulsating coordinates.
\begin{equation*}
  x = \frac{1}{r_{EM}}X,\quad y = \frac{1}{r_{EM}}Y,\quad z=\frac{1}{r_{EM}}Z
\end{equation*}
Let us focus on just the $x$ coordinate for now. Taking the derivative with respect to true anomaly yields
\begin{equation*}
\begin{aligned}
  x' &= \left(\frac 1 {r_{EM}}\right)'X + \frac 1{r_{EM}}X'\\
     &= \frac{\textrm d}{\textrm d t}\left(\frac 1{r_{EM}}\right)\frac{1}{\dot f}X + \frac 1{\dot fr_{EM}}\dot X
\end{aligned}
\end{equation*}
where 
\begin{equation*}
\begin{aligned}
  \frac{\textrm d}{\textrm d t}\left(\frac 1{r_{EM}}\right) &= \frac{\textrm d}{\textrm d t}\left(\frac 1{|\mathbf{r}_{EM}|}\right) = -\frac{\mathbf r_{EM} \cdot \mathbf v_{EM}}{r_{EM}^3}.
\end{aligned}
\end{equation*}
Thus, the rotating state can be transformed into the $f$-dependent rotating-pulsating state with the following two equations.
\begin{equation}\label{eq:rotpulse}
\begin{aligned}
  {}^P\mathbf r &= \frac{1}{r_{EM}}{}^R\mathbf r\\
  {}^P\mathbf v &= -\left(\frac{\mathbf r_{EM}\cdot \mathbf v_{EM}}{\dot fr_{EM}^3}\right){}^R\mathbf r + \left(\frac 1{\dot fr_{EM}}\right){}^R\mathbf v
\end{aligned}
\end{equation}
In the CR3BP, the true anomaly is equivalent to the mean anomaly
\begin{equation*}
  f=nt,
\end{equation*}
meaning that $\dot{f}=n$ where
\begin{equation}
  n = \sqrt{\frac{G(m_S+m_E+m_M)}{r_{EM}^3}}.
\end{equation}
Therefore, $\dot{f}$ can be replaced with $n$ in \eqref{eq:rotpulse} and the notation $()'$ will then denote the derivative with respect to non-dimensional time.

\subsection{True anomaly}
When working with the ER3BP, it is also necessary to know the true anomaly associated with a given state.
To find the true anomaly of the Earth-Moon barycenter, the eccentricity vector must first be calculated with \eqref{eq:evec}.
\begin{equation}\label{eq:evec}
  \mathbf e = \frac{\mathbf r_{EM}\times \mathbf v_{EM}}{\hat\mu} - \frac{\mathbf r_{EM}}{r_{EM}}
\end{equation}
Where $\hat \mu$ is the two-body mass parameter of the system, not to be confused with the mass parameter of the three-body problem, $\mu$. The value of $\hat \mu$ can be calculated using the constants defined earlier in \eqref{eq:Gmvals}..
\begin{equation}
\hat \mu = G(m_S+m_E+m_M)
\end{equation}
The eccentricity is then just the magnitude of the eccentricity vector.
\begin{equation}
e = |\mathbf e|
\end{equation}
Finally, the true anomaly $f$ corresponding to the given $\mathbf r_{EM}$ and $\mathbf v_{EM}$ vectors can be calculated with the following equation.
\begin{equation}
  f = \left\{\begin{matrix}\cos^{-1}\left(\frac{\mathbf e\cdot\mathbf r_{EM}}{er_{EM}}\right) & \frac{\mathbf r_{EM}\cdot \mathbf v_{EM}}{r_{EM}} \geq 0 \\ 2\pi-\cos^{-1}\left(\frac{\mathbf e\cdot\mathbf r_{EM}}{er_{EM}}\right) & \frac{\mathbf r_{EM} \cdot\mathbf v_{EM}}{r_{EM}} < 0 
\end{matrix} \right.
\end{equation}
\bibliography{sample}

@article{paez2021transits,
  title={Transits close to the Lagrangian solutions L 1, L 2 in the elliptic restricted three-body problem},
  author={Paez, Roc{\'\i}o I and Guzzo, Massimiliano},
  journal={Nonlinearity},
  volume={34},
  number={9},
  pages={6417},
  year={2021},
  publisher={IOP Publishing}
}

@article{paez2022semi,
  title={On the semi-analytical construction of halo orbits and halo tubes in the elliptic restricted three-body problem},
  author={Paez, Roc{\'\i}o I and Guzzo, Massimiliano},
  journal={Physica D: Nonlinear Phenomena},
  volume={439},
  pages={133402},
  year={2022},
  publisher={Elsevier}
}

@article{jorba1998numerical,
  title={Numerical computation of normal forms around some periodic orbits of the restricted three-body problem},
  author={Jorba, Angel and Villanueva, Jordi},
  journal={Physica D: Nonlinear Phenomena},
  volume={114},
  number={3-4},
  pages={197--229},
  year={1998},
  publisher={Elsevier}
}

@article{jorba1999methodology,
  title={A methodology for the numerical computation of normal forms, centre manifolds and first integrals of Hamiltonian systems},
  author={Jorba, Angel},
  journal={Experimental Mathematics},
  volume={8},
  number={2},
  pages={155--195},
  year={1999},
  publisher={Taylor \& Francis}
}

@article{jorba2020vicinity,
  title={The vicinity of the Earth--Moon L 1 point in the bicircular problem},
  author={Jorba, {\`A}ngel and Jorba-Cusc{\'o}, Marc and Rosales, Jos{\'e} J},
  journal={Celestial Mechanics and Dynamical Astronomy},
  volume={132},
  number={2},
  pages={11},
  year={2020},
  publisher={Springer}
}

@article{gabern2001restricted,
  title={A restricted four-body model for the dynamics near the Lagrangian points of the Sun-Jupiter system},
  author={Gabern, Frederic and Jorba, Angel},
  journal={Discrete and Continuous Dynamical Systems Series B},
  volume={1},
  number={2},
  pages={143--182},
  year={2001},
  publisher={AIMS PRESS}
}

@book{gomez2001dynamics,
  title={Dynamics and Mission Design Near Libration Points, Vol IV: Advanced Methods for Triangular Points},
  author={G{\'o}mez, Gerard and Jorba, Angel and Masdemont, Josep J and Simo, Carles},
  volume={5},
  year={2001},
  publisher={world scientific}
}

@article{le2017invariant,
  title={Invariant manifolds of a non-autonomous quasi-bicircular problem computed via the parameterization method},
  author={Le Bihan, Bastien and Masdemont, JJ and G{\'o}mez, G and Lizy-Destrez, Stephanie},
  journal={Nonlinearity},
  volume={30},
  number={8},
  pages={3040},
  year={2017},
  publisher={IOP Publishing}
}

@phdthesis{andreu1998quasi,
  title={The quasi-bicircular problem},
  author={Andreu, MA},
  year={1998},
  school={PhD thesis, Univ. Barcelona}
}

@article{andreu2002dynamics,
  title={Dynamics in the center manifold around L2 in the quasi-bicircular problem},
  author={Andreu, MA},
  journal={Celestial Mechanics and Dynamical Astronomy},
  volume={84},
  number={2},
  pages={105--133},
  year={2002},
  publisher={Springer}
}

@article{rosales2020effect,
  title={On the effect of the Sun's gravity around the Earth-Moon L1 and L2 libration points},
  author={Rosales de C{\'a}ceres, Jos{\'e} J},
  year={2020},
  publisher={Universitat de Barcelona}
}

@article{peterson2023vicinity,
  title={The vicinity of Earth--Moon L1 and L2 in the Hill restricted 4-body problem},
  author={Peterson, Luke T and Rosales, Jos{\'e} J and Scheeres, Daniel J},
  journal={Physica D: Nonlinear Phenomena},
  volume={455},
  pages={133889},
  year={2023},
  publisher={Elsevier}
}

@article{peterson2023local,
  title={Local orbital elements for the circular restricted three-body problem},
  author={Peterson, Luke T and Scheeres, Daniel J},
  journal={Journal of Guidance, Control, and Dynamics},
  volume={46},
  number={12},
  pages={2275--2289},
  year={2023},
  publisher={American Institute of Aeronautics and Astronautics}
}

@article{peterson2024dynamics,
  title={Dynamics around the Earth--Moon triangular points in the Hill restricted 4-body problem},
  author={Peterson, Luke T and Brown, Gavin and Jorba, {\`A}ngel and Scheeres, Daniel},
  journal={Celestial Mechanics and Dynamical Astronomy},
  volume={136},
  number={4},
  pages={31},
  year={2024},
  publisher={Springer}
}

@article{schwab2024characterizing,
  title={Characterizing Accuracy of Normal Forms to Study Trajectories in Cislunar Space},
  author={Schwab, David and Eapen, Roshan and Singla, Puneet},
  journal={The Journal of the Astronautical Sciences},
  volume={71},
  number={2},
  pages={16},
  year={2024},
  publisher={Springer}
}

@phdthesis{schwabCislunarTransportCharacterization2024,
  title = {Cislunar {{Transport Characterization}} for {{Space Situational Awareness}}},
  author = {Schwab, David},
  year = {2024},
  month = may,
  address = {University Park, PA},
  school = {The Pennsylvania State University}
}

@article{zhao2025lie,
  title={Lie-series transformations and applications to construction of analytical solution},
  author={Zhao, Shunjing and Lei, Hanlun},
  journal={Nonlinear Dynamics},
  volume={113},
  number={3},
  pages={2183--2198},
  year={2025},
  publisher={Springer}
}

@article{richardson1980analytic,
  title={Analytic construction of periodic orbits about the collinear points},
  author={Richardson, David L},
  journal={Celestial mechanics},
  volume={22},
  number={3},
  pages={241--253},
  year={1980},
  publisher={Springer}
}

@article{celletti2024dynamics,
  title={The dynamics around the collinear points of the elliptic three-body problem: A normal form approach},
  author={Celletti, Alessandra and Lhotka, Christoph and Pucacco, Giuseppe},
  journal={Physica D: Nonlinear Phenomena},
  volume={468},
  pages={134302},
  year={2024},
  publisher={Elsevier}
}

@inproceedings{dichmann2014stationkeeping,
  title={Stationkeeping Monte Carlo simulation for the James Webb space telescope},
  author={Dichmann, Donald J and Alberding, Cassandra M and Yu, Wayne H},
  booktitle={International Symposium on Space Flight Dynamics 2014},
  number={GSFC-E-DAA-TN14095},
  year={2014}
}

@book{szebehely2012theory,
  title={Theory of orbit: The restricted problem of three Bodies},
  author={Szebehely, Victory},
  year={2012},
  publisher={Elsevier}
}

@book{cabral2023normal,
  title={Normal Forms and Stability of Hamiltonian Systems},
  author={Cabral, Hildeberto E and Dias, L{\'u}cia Brand{\~a}o},
  year={2023},
  publisher={Springer}
}

@unpublished{hunsbergerschwab2025hawaii,
  author       = {Carson Hunsberger and David Schwab and Roshan Eapen and Puneet Singla},
  title        = {Comparing Normal Form Representations for Station-Keeping near Cislunar Libration Points},
  note         = {Preprint, submitted to AAS/AIAA Space Flight Mechanics Meeting},
  year         = {2025},
  url          = {https://www.xcdsystem.com/aas/program/riGLvh6/index.cfm?pgid=5425},
}

@article{park2021jpl,
  title={The JPL planetary and lunar ephemerides DE440 and DE441},
  author={Park, Ryan S and Folkner, William M and Williams, James G and Boggs, Dale H},
  journal={The Astronomical Journal},
  volume={161},
  number={3},
  pages={105},
  year={2021},
  publisher={IOP Publishing}
}
\clearpage
\begin{figure}[htb!]
    \centering
    \subfigure[Earth-Moon ER3BP normal form Lyapunov trajectory ($I_2=0.2$)\label{subfig:EMLyapunovrotating}]{
        {\includegraphics[width=0.35\textwidth]{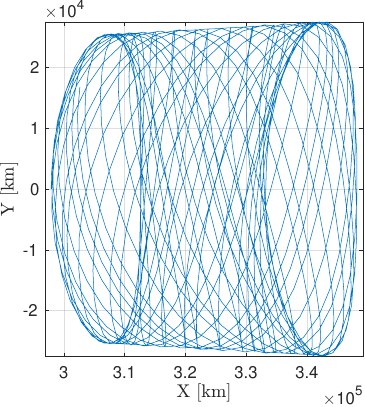}}
        }\hfill
    \subfigure[Sun-Earth ER3BP normal form Lyapunov trajectory ($I_2=0.2$)\label{subfig:SELyapunovrotating}]{
        {\includegraphics[width=0.45\textwidth]{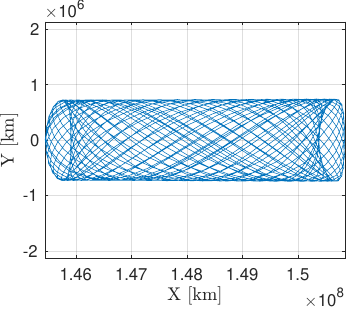}}
        }\\
    \subfigure[Earth-Moon ER3BP normal form vertical trajectory ($\hat I_3=0.3$)\label{subfig:EMverticalrotating}]{
        {\includegraphics[width=0.38\textwidth]{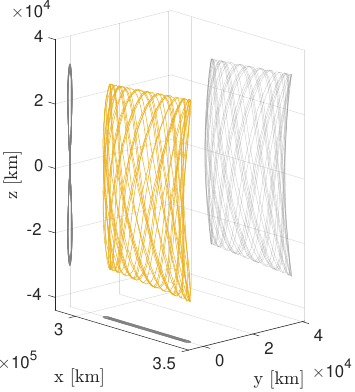}}
        }\hfill
    \subfigure[Sun-Earth ER3BP normal form vertical trajectory ($\hat I_3=0.3$)\label{subfig:SEverticalrotating}]{
        {\includegraphics[width=0.58\textwidth]{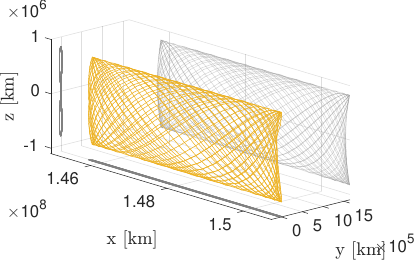}}
        }\\
    \subfigure[Earth-Moon ER3BP normal form halo trajectory $(\hat I_2,\hat I_3,\theta_2)\approx (0.2116,0.3,-\pi/2)$\label{subfig:EMhalorotating}]{
        {\includegraphics[width=0.44\textwidth]{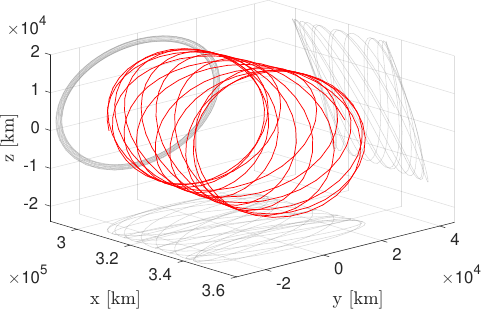}}
        }\hfill
    \subfigure[Earth-Moon ER3BP normal form halo trajectory $(\hat I_2,\hat I_3,\theta_2)\approx (0.2246,0.3,-\pi/2)$\label{subfig:SEhalorotating}]{
        {\includegraphics[width=0.52\textwidth]{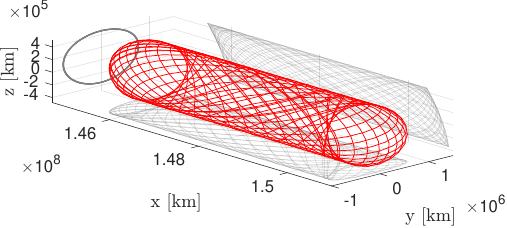}}
        }\\
        \caption{Comparison between Earth-Moon and Sun-Earth center manifold trajectories in the non-pulsating frame.\label{fig:EMvsSE}}
\end{figure}
\begin{figure}[htb!]
    \centering
    \subfigure[Earth-Moon ER3BP normal form Lissajous trajectory $(\hat I_2(0),\hat I_3,\theta_2(0))\approx (0.1,0.107,0)$\label{subfig:EMLissajousrotating}]{
        {\includegraphics[width=0.44\textwidth]{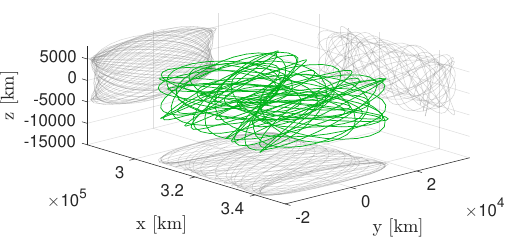}}
        }\hfill
    \subfigure[Sun-Earth ER3BP normal form Lissajous trajectory $(\hat I_2(0),\hat I_3,\theta_2(0))\approx (0.1,0.107,0)$\label{subfig:SELissajousrotating}]{
        {\includegraphics[width=0.50\textwidth]{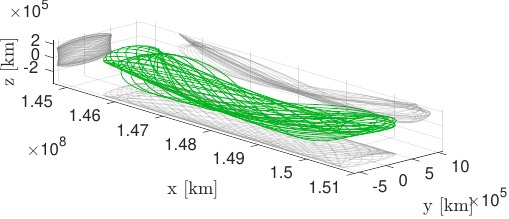}}
        }\\
    \subfigure[Earth-Moon ER3BP normal form quasihalo trajectory $(\hat I_2(0),\hat I_3,\theta_2(0))\approx (0.2,0.3,-\pi/2)$\label{subfig:EMquasihalorotating}]{
        {\includegraphics[width=0.45\textwidth]{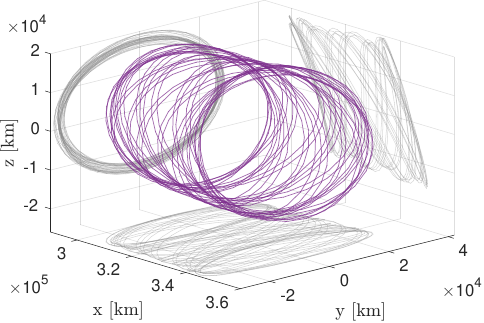}}
        }\hfill
    \subfigure[Sun-Earth ER3BP normal form quasihalo trajectory $(\hat I_2(0),\hat I_3,\theta_2(0))\approx (0.2,0.3,-\pi/2)$\label{subfig:SEquasihalorotating}]{
        {\includegraphics[width=0.52\textwidth]{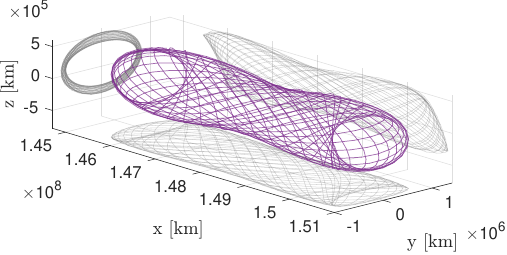}}
        }\\
        \caption{Comparison between Earth-Moon and Sun-Earth center manifold trajectories in the non-pulsating frame.\label{fig:EMvsSE2}}
\end{figure}

\clearpage
\section*{Tables of Action-Angle Hamiltonian Coefficients}
\begin{table}[htb!]
  \centering
  \caption{Birkhoff $L_1$ Earth-Moon action-angle Hamiltonian coefficients ($\mu=0.012150585$ and $e=0.0549006$).\label{tab:L1Birkhoffcoeffs}}
  \begin{minipage}[t][10cm][t]{0.25\linewidth}
    \vspace{0pt}
\begin{tabular}{|r|c|c|c|}
\hline
$H_1(-,\mathbf I)$ & $I_1$ & $I_2$ & $I_3$ \\
\hline
2.9341 & 1 & 0 & 0 \\
2.3355 & 0 & 1 & 0 \\
2.2699 & 0 & 0 & 1 \\
\hline
\end{tabular}\\
\begin{tabular}{|r|c|c|c|}
\hline
$H_2(-,\mathbf I)$ & $I_1$ & $I_2$ & $I_3$ \\
\hline
-0.2194 & 2 & 0 & 0 \\
-0.7538 & 1 & 1 & 0 \\
-0.6893 & 1 & 0 & 1 \\
-0.1621 & 0 & 2 & 0 \\
-0.0726 & 0 & 1 & 1 \\
-0.1449 & 0 & 0 & 2 \\
\hline
\end{tabular}\\
\begin{tabular}{|r|c|c|c|}
\hline
$H_3(-,\mathbf I)$ & $I_1$ & $I_2$ & $I_3$ \\
\hline
-0.0287 & 3 & 0 & 0 \\
-0.1113 & 2 & 1 & 0 \\
-0.0745 & 2 & 0 & 1 \\
-0.0255 & 1 & 2 & 0 \\
-0.1180 & 1 & 1 & 1 \\
-0.0077 & 1 & 0 & 2 \\
-0.0133 & 0 & 3 & 0 \\
0.4115 & 0 & 2 & 1 \\
-0.4167 & 0 & 1 & 2 \\
-0.0084 & 0 & 0 & 3 \\
\hline
\end{tabular}

  \end{minipage}
  \begin{minipage}[t][10cm][t]{0.25\linewidth}
    \vspace{0pt}
\begin{tabular}{|r|c|c|c|}
\hline
$H_4(-,\mathbf I)$ & $I_1$ & $I_2$ & $I_3$ \\
\hline
-0.0068 & 4 & 0 & 0 \\
-0.0316 & 3 & 1 & 0 \\
-0.0128 & 3 & 0 & 1 \\
0.0252 & 2 & 2 & 0 \\
0.0568 & 2 & 1 & 1 \\
0.0337 & 2 & 0 & 2 \\
0.0035 & 1 & 3 & 0 \\
0.0824 & 1 & 2 & 1 \\
 -0.1135& 1 & 1 & 2 \\
 0.0080& 1 & 0 & 3 \\
-0.0012 & 0 & 4 & 0 \\
1.6997 & 0 & 3 & 1 \\
-4.4568 & 0 & 2 & 2 \\
1.2754 & 0 & 1 & 3 \\
-0.0002 & 0 & 0 & 4 \\
\hline
\end{tabular}
\vfill
  \end{minipage}
  \begin{minipage}[t]{0.25\linewidth}
    \vspace{0pt}
\begin{tabular}{|r|c|c|c|}
\hline
$H_5(-,\mathbf I)$ & $I_1$ & $I_2$ & $I_3$ \\
\hline
-0.0018 & 5 & 0 & 0 \\
-0.0105 & 4 & 1 & 0 \\
-0.0012 & 4 & 0 & 1 \\
0.0188  & 3 & 2 & 0 \\
0.0341  & 3 & 1 & 1 \\
0.0205  & 3 & 0 & 2 \\
0.0054  & 2 & 3 & 0 \\
0.4264  & 2 & 2 & 1 \\
 -0.3354& 2 & 1 & 2 \\
 0.0083 & 2 & 0 & 3 \\
0.0023  & 1 & 4 & 0 \\
1.4808  & 1 & 3 & 1 \\
-3.8650 & 1 & 2 & 2 \\
1.1023  & 1 & 1 & 3 \\
0.0031 & 1 & 0 & 4 \\
 0.0001 & 0 & 5 & 0 \\
8.0880  & 0 & 4 & 1 \\
-39.7066  & 0 & 3 & 2 \\
34.3424 & 0 & 2 & 3 \\
-5.4009  & 0 & 1 & 4 \\
0.0003 & 0 & 0 & 5 \\
\hline
\end{tabular}
  \end{minipage}
\end{table}

\clearpage
\begin{table}[htb!]
  \centering
  \caption{Resonant $L_1$ Earth-Moon action-angle Hamiltonian coefficients ($\mu=0.012150585$ and $e=0.0549006$).\label{tab:L1EMresonant}}
  \begin{minipage}[t][10cm][t]{0.4\linewidth}
    \vspace{0pt}
\begin{tabular}{|r|c|c|c|c|}
\hline
$H_1(\theta_2,\mathbf I)$ & $I_1$ & $I_2$ & $I_3$ & $\theta_2$\\
\hline
2.9341 & 1 & 0 & 0 &  \\
0.0655 & 0 & 1 & 0 & \\
2.2699 & 0 & 0 & 1 & \\
\hline
\end{tabular}\\
\begin{tabular}{|r|c|c|c|c|}
\hline
$H_2(\theta_2,\mathbf I)$ & $I_1$ & $I_2$ & $I_3$ & $\theta_2$\\
\hline
-0.2194 & 2 & 0 & 0&\\
-0.0645 & 1 & 1 & 0& \\
-0.6893 & 1 & 0 & 1& \\
-0.2344 & 0 & 2 & 0& \\
0.2331  & 0 & 2 & 0  & $\cos(2\theta_2)$\\
0.2172  & 0 & 1 & 1& \\
-0.2331 & 0 & 1 & 1 & $\cos(2\theta_2)$ \\
-0.1449 & 0 & 0 & 2& \\
\hline
\end{tabular}\\
\begin{tabular}{|r|c|c|c|c|}
\hline
$H_3(\theta_2,\mathbf I)$ & $I_1$ & $I_2$ & $I_3$ & $\theta_2$ \\
\hline
-0.0287 & 3 & 0 & 0& \\
-0.0368 & 2 & 1 & 0& \\
-0.0745 & 2 & 0 & 1& \\
0.0848  & 1 & 2 & 0& \\
-0.0849 & 1 & 2 & 0  & $\cos(2\theta_2)$\\
-0.1026  & 1 & 1 & 1& \\
0.0849 & 1 & 1 & 1  & $\cos(2\theta_2)$\\
-0.0077 & 1 & 0 & 2& \\
-0.0042 & 0 & 3 & 0& \\
 0.0042 & 0 & 3 & 0  & $\cos(2\theta_2)$\\
 -0.0236  & 0 & 2 & 1& \\
0.0234 & 0 & 2 & 1  & $\cos(2\theta_2)$\\
0.0230  & 0 & 1 & 2& \\
-0.0276 & 0 & 1 & 2  & $\cos(2\theta_2)$\\
-0.0084 & 0 & 0 & 3& \\
\hline
\end{tabular}

  \end{minipage}
  \begin{minipage}[t]{0.4\linewidth}
    \vspace{0pt}
\begin{tabular}{|r|c|c|c|c|}
\hline
$H_4(\theta_2,\mathbf I)$ & $I_1$ & $I_2$ & $I_3$ & $\theta_2$\\
\hline
-0.0068 & 4 & 0 & 0& \\
-0.0188 & 3 & 1 & 0& \\
-0.0128 & 3 & 0 & 1& \\
0.0022 & 2 & 2 & 0& \\
-0.0038  & 2 & 2 & 0  & $\cos(2\theta_2)$\\
-0.0107  & 2 & 1 & 1& \\
0.0038 & 2 & 1 & 1  & $\cos(2\theta_2)$\\
0.0337 & 2 & 0 & 2& \\
0.0112 & 1 & 3 & 0& \\
-0.0111 & 1 & 3 & 0  & $\cos(2\theta_2)$\\ 
0.0161 & 1 & 2 & 1& \\
-0.0173 & 1 & 2 & 1  & $\cos(2\theta_2)$\\ 
-0.0318  & 1 & 1 & 2& \\
0.0284  & 1 & 1 & 2  & $\cos(2\theta_2)$\\ 
0.0080 & 1 & 0 & 3& \\
0.0136 & 0 & 4 & 0& \\
-0.0181 & 0 & 4 & 0  & $\cos(2\theta_2)$\\
0.0046 & 0 & 4 & 0  & $\cos(4\theta_2)$\\
-0.0250 & 0 & 3 & 1& \\
0.0342 & 0 & 3 & 1  & $\cos(2\theta_2)$\\
-0.0092  & 0 & 3 & 1  & $\cos(4\theta_2)$\\
0.0095  & 0 & 2 & 2& \\
-0.0143 & 0 & 2 & 2  & $\cos(2\theta_2)$\\
0.0046 & 0 & 2 & 2  & $\cos(4\theta_2)$\\
0.0009  & 0 & 1 & 3& \\
-0.0018 & 0 & 1 & 3  & $\cos(2\theta_2)$\\
-0.0002 & 0 & 0 & 4& \\
\hline
\end{tabular}
  \end{minipage}
\end{table}
\clearpage

\begin{table}[htb!]
  \centering
  \caption{Resonant $L_2$ Sun-EM barycenter action-angle Hamiltonian coefficients with SRP ($\mu=3.040423404760033\times10^{-6}$, $e=0.01671022$, $\Theta=2.979677739179855\times 10^{-5}$).\label{tab:resonantL2srp}}
  \begin{minipage}[t][10cm][t]{0.4\linewidth}
    \vspace{0pt}
\begin{tabular}{|r|c|c|c|c|}
\hline
$H_1(\theta_2,\mathbf I)$ & $I_1$ & $I_2$ & $I_3$ & $\theta_2$\\
\hline
2.4857 & 1 & 0 & 0 & \\
0.0719 & 0 & 1 & 0 & \\
1.9859 & 0 & 0 & 1 & \\
\hline
\end{tabular}\\
\begin{tabular}{|r|c|c|c|c|}
\hline
$H_2(\theta_2,\mathbf I)$ & $I_1$ & $I_2$ & $I_3$ & $\theta_2$\\
\hline
-0.1126 & 2 & 0 & 0 & \\
-0.0665 & 1 & 1 & 0  & \\
-0.4586 & 1 & 0 & 1  & \\
-0.2021 & 0 & 2 & 0 & \\
0.2009  & 0 & 2 & 0  & $\cos(2\theta_2)$\\
0.1841  & 0 & 1 & 1  & \\
-0.2009 & 0 & 1 & 1 & $\cos(2\theta_2)$ \\
-0.0742 & 0 & 0 & 2  & \\
\hline
\end{tabular}\\
\begin{tabular}{|r|c|c|c|c|}
\hline
$H_3(\theta_2,\mathbf I)$ & $I_1$ & $I_2$ & $I_3$ & $\theta_2$ \\
\hline
-0.0297 & 3 & 0 & 0 & \\
-0.0386 & 2 & 1 & 0 & \\
-0.0950 & 2 & 0 & 1 & \\
0.0698  & 1 & 2 & 0 & \\
-0.0709 & 1 & 2 & 0  & $\cos(2\theta_2)$\\
-0.0910  & 1 & 1 & 1 & \\
0.0709 & 1 & 1 & 1  & $\cos(2\theta_2)$\\
-0.0426 & 1 & 0 & 2 & \\
-0.0033 & 0 & 3 & 0 & \\
 0.0032 & 0 & 3 & 0  & $\cos(2\theta_2)$\\
 -0.0393  & 0 & 2 & 1 & \\
0.0388 & 0 & 2 & 1  & $\cos(2\theta_2)$\\
0.0369  & 0 & 1 & 2 & \\
-0.0420 & 0 & 1 & 2  & $\cos(2\theta_2)$\\
-0.0121 & 0 & 0 & 3 & \\
\hline
\end{tabular}

  \end{minipage}
  \begin{minipage}[t]{0.4\linewidth}
    \vspace{0pt}
\begin{tabular}{|r|c|c|c|c|}
\hline
$H_4(\theta_2,\mathbf I)$ & $I_1$ & $I_2$ & $I_3$ & $\theta_2$\\
\hline
-0.0081 & 4 & 0 & 0 & \\
-0.0204 & 3 & 1 & 0 & \\
-0.0190 & 3 & 0 & 1 & \\
0.0098 & 2 & 2 & 0 & \\
-0.0138  & 2 & 2 & 0  & $\cos(2\theta_2)$\\
-0.0221  & 2 & 1 & 1 & \\
0.0138 & 2 & 1 & 1  & $\cos(2\theta_2)$\\
0.0206 & 2 & 0 & 2 & \\
0.0131 & 1 & 3 & 0 & \\
-0.0130 & 1 & 3 & 0  & $\cos(2\theta_2)$\\ 
0.0263 & 1 & 2 & 1 & \\
-0.0288 & 1 & 2 & 1  & $\cos(2\theta_2)$\\ 
-0.0454  & 1 & 1 & 2 & \\
0.0418  & 1 & 1 & 2  & $\cos(2\theta_2)$\\ 
0.0033 & 1 & 0 & 3 & \\
0.0159 & 0 & 4 & 0 & \\
-0.0214 & 0 & 4 & 0  & $\cos(2\theta_2)$\\
0.0054 & 0 & 4 & 0  & $\cos(4\theta_2)$\\
-0.0303 & 0 & 3 & 1 & \\
0.0411 & 0 & 3 & 1  & $\cos(2\theta_2)$\\
-0.0109  & 0 & 3 & 1  & $\cos(4\theta_2)$\\
0.0086  & 0 & 2 & 2 & \\
-0.0144 & 0 & 2 & 2  & $\cos(2\theta_2)$\\
0.0054 & 0 & 2 & 2  & $\cos(4\theta_2)$\\
0.0043  & 0 & 1 & 3 & \\
-0.0053 & 0 & 1 & 3  & $\cos(2\theta_2)$\\
-0.0013 & 0 & 0 & 4 & \\
\hline
\end{tabular}
  \end{minipage}
\end{table}

\end{document}